\documentclass[english,pra,reprint, aps, showkeys, showpacs, superscriptaddress]{revtex4-1}
\usepackage[LGR,T1]{fontenc}
\usepackage[latin9]{inputenc}
\usepackage{float}
\usepackage{amsmath}
\usepackage{amssymb}
\usepackage{stackrel}
\usepackage{graphicx}
\usepackage{color,xcolor}
\makeatletter

\pdfpageheight\paperheight
\pdfpagewidth\paperwidth

\DeclareRobustCommand{\greektext}{%
	\fontencoding{LGR}\selectfont\def\encodingdefault{LGR}}
\DeclareRobustCommand{\textgreek}[1]{\leavevmode{\greektext #1}}
\ProvideTextCommand{\~}{LGR}[1]{\char126#1}

\usepackage[colorlinks,citecolor=blue,urlcolor=blue]{hyperref}

\usepackage{ulem} 

\makeatother

\usepackage{babel}
\begin{document}
\title{Arbitrary entangled state transfer via a topological qubit chain}
\author{Chong Wang}
\affiliation{School of integrated circuits, Tsinghua University, Beijing, 100084,
China}
\affiliation{Frontier Science Center for Quantum Information, Beijing, China}
\author{Linhu Li}
\affiliation{Guangdong Provincial Key Laboratory of Quantum Metrology and Sensing
\& School of Physics and Astronomy, Sun Yat-Sen University (Zhuhai
Campus), Zhuhai 519082, China}
\author{Jiangbin Gong}
\email{phygj@nus.edu.sg}
\affiliation{Department of Physics, National University of Singapore, Singapore 117542, Singapore}
\author{Yu-xi Liu}
\email{yuxiliu@mail.tsinghua.edu.cn}
\affiliation{School of integrated circuits, Tsinghua University, Beijing, 100084,
China}
\affiliation{Frontier Science Center for Quantum Information, Beijing, China}
\date{\today}
\begin{abstract}
Quantum state transfer is one of the basic tasks in quantum information processing. We here propose a theoretical approach to realize arbitrary entangled state transfer through a qubit chain, which is a class of extended Su-Schrieffer-Heeger models and accommodates multiple topological edge states separated from the bulk states. We show that an arbitrary entangled state, from $2$-qubit to $\mathcal{N}$-qubit, can be encoded in the corresponding edge states, and then adiabatically transferred from one end to the other of the chain. The dynamical phase differences resulting from the time evolutions of different edge states can be eliminated by properly choosing evolution time. Our approach is robust against both the qubit-qubit coupling disorder and the evolution time disorder. For the concreteness of discussions, we assume that such a chain is constructed by an experimentally feasible superconducting qubit system, meanwhile, our proposal can also be applied to other systems.
\end{abstract}
\maketitle

\section{introduction}
There are many possible platforms to realize quantum computation~\citep{steane1998quantum,preskill2018quantum,bennett2000quantum},
e.g., cold atoms~\citep{weiss2017quantum,negretti2011quantum,briegel2000quantum,garcia2005quantum},
trapped ions~\citep{garcia2005quantum,haffner2008quantum,benhelm2008towards,bruzewicz2019trapped,cirac2000scalable,lanyon2013measurement,garcia2003speed,pachos2002quantum,feng2002quantum},
and superconducting quantum circuits~\citep{jeffrey2014fast,gambetta2017building,brecht2016multilayer,devoret2004superconducting,you2006superconducting,yan2019strongly,song201710,neill2018blueprint,huang2020superconducting,krantz2019quantum,you2003quantum,osborn2007frequency,pechal2018superconducting}.
On these platforms, quantum state transfer (QST) in a controllable way~\citep{bennett2000quantum,bouwmeester2000physics}
is one of crucial requirements.
Though long-distance quantum communication has been widely achieved in
optical fibers and free-space~\citep{zhu2017experimental,liu2019energy,huo2018deterministic,chen2021integrated,liao2018satellite}, it is still very important to find a promising way for transferring quantum states through solid-state devices or condensed matter~\citep{bose2003quantum,christandl2004perfect,balachandran2008adiabatic,song2005quantum,yung2006quantum,yung2005perfect,wu2009perfect,wu2009universal}.
A number of QST protocols have been proposed for different solid-state medium~\citep{bienfait2019phonon,sete2015high,vermersch2017quantum,he2017quantum,maring2017photonic,northup2014quantum,cirac1997quantum,shi2005quantum,zwick2011robustness,kandel2021adiabatic}. In recent years, QST via a spin chain has attracted extensive attentions~\citep{bose2003quantum,christandl2004perfect,balachandran2008adiabatic}
and many technologies have already been developed ~\citep{shi2005quantum,zwick2011robustness,kandel2021adiabatic,PhysRevLett.106.040505,bose2007quantum,mei2018robust,longhi2019topological,d2020fast,wang2018dynamical}.

Perfect QST can be realized through well designed spin chain with invariable couplings~\citep{Zhang2005Simulation,Petrosyan2010State,Gualdi2008Perfect}. Meanwhile,
it can also be realized by precisely modulating the spin-spin couplings~\citep{Wang2020Almost,lyakhov2006use,zwick2014optimized,benjamin2001quantum}.
For example, quantum states can be transferred by simply applying a sequence of SWAP operations implemented by
\textgreek{p} pulses between the pairs of nearest neighboring sites~\citep{Petrosyan2010State}, which needs only the spin-spin couplings to be switched on and off periodically~\citep{benjamin2003quantum}. Nevertheless,
these known methods require the accurate design of the system Hamiltonian, thus are usually
less robust against disorders and imperfections in large-scale implementations.
To overcome this issue, adiabatic QST protocols have been widely studied~\citep{balachandran2008adiabatic,tan2020high,demirplak2003adiabatic,gong2004complete,gong2004adiabatic,eckert2007efficient,ho2012quantized,tian2012adiabatic,agundez2017superadiabatic,sandberg2016efficient}, as the QST exploiting the adiabatic theorem~\citep{messiah1962quantum,berry1984quantal} is independent of the protocol operation details so long as
the evolution of the system is slow enough.

Our QST method here is based on the spectral and dynamical features of a topological qubit chain. For topological insulators, robust conducting edge states are guaranteed by the nontrivial topology of bulk bands~\citep{hasan2010colloquium,qi2011topological,asboth2016short,luo2019advanced}.
These edge states are insensitive to smooth changes in the system parameters unless
a topological phase transition occurs~\citep{hasan2010colloquium,qi2011topological}.
Such robustness based on topological protection provides topological quantum
systems a great potential for quantum information and quantum computing~\citep{nayak2008non,freedman2003topological,stern2013topological,bomantara2018simulation,bomantara2018quantum,bomantara2020measurement}.
The Su-Schrieffer-Heeger (SSH) chain~\citep{su1979solitons} is the simplest model of the topological insulators and can  be realized by a qubit chain with staggered couplings when the qubit chain is restricted to the subspace of single excitation.

Building on the concepts of topological edge states and adiabatic QST via a qubit chain,
protocols for efficient QST~\citep{yao2013topologically,mei2018robust,palaiodimopoulos2021fast,longhi2019topological,d2020fast,wang2018dynamical}
have also been proposed, with more robustness to disorder due to the underlying topological protection. With specific design of topological qubit chains, quantum
states can be encoded in edge states of the systems, and transferred from one end of the chain to the other
by adiabatically altering couplings between qubits~\citep{mei2018robust,palaiodimopoulos2021fast,longhi2019topological,d2020fast}.
However, most of these available proposals focus on single qubit state transfer, and multi-qubit state transfer with arbitrary entanglement is still a challenging task. Recently, a more advanced protocol for QST via the so-called Floquet topological edge modes was proposed~\citep{tan2020high}. In this protocol, some entangled states are encoded in the edge states of quasienergy zero and $\pi$ modes, and the high-fidelity transfer of entangled states can be achieved over a long distance.
However, this method requires additional dynamical modulation on the time-periodic couplings between the qubits and may pose new experimental challenges.

In our method, arbitrary entangled state can be encoded in topological edge states, which are supported by a generalized SSH chain. As the parameters of the system are slowly altered, the edge states can be adiabatically transferred from one end of the chain to the other. Thus the entangled state can be transferred along this adiabatic passage. It is well known that the adiabatic evolution leads to two phases, i.e, geometric phase and dynamical phase. The geometric
phase can be easily wiped out with a chosen gauge in non-closed adiabatic
evolution~\citep{berry1984quantal}, and the dynamical phase can be eliminated when the evolution time
is carefully chosen. Thus, the quantum phases between different components encoding an entangled state can be well-controlled or recovered, and the entangled state can be truly transferred.

For the concreteness of discussions, we assume that the qubit chain is formed by the superconducting quantum circuits, which have been developing rapidly~\citep{jeffrey2014fast,gambetta2017building,brecht2016multilayer}.
In particular, superconducting qubit chains have been widely studied for simulating many-body quantum physics~\citep{houck2012chip,paraoanu2014recent,roushan2017spectroscopic}.
There are several advantages to use superconducting
qubits as quantum simulators. First,  superconducting
qubits are highly coherent with long coherence time ($10\sim100\mu s$)~\citep{PhysRevLett.107.240501,novikov2016raman}. Second, most of the parameters of superconducting
qubits are highly controllable~\citep{chen2014qubit,geller2015tunable,barends2015digital,reuther2010two,baust2015tunable},
thus we can perform rather arbitrary operation on the systems.
Third, the superconducting quantum circuits have high scalability and designability.
One quantum chip can support a large number of controllable qubits with different connecting manners~\citep{arute2019quantum}.
These advantages make superconducting quantum circuits one of the best platforms to perform quantum simulation and quantum computing. Moreover, the topological chain constructed by superconducting qubits has already been experimentally demonstrated~\citep{cai2019observation}. Thus, for the state-of-the-art, superconducting qubit circuits are very suitable to construct such a chain for QST. The system
parameters considered in this work are all based on the existing experiments of Xmon qubits~\citep{chen2014qubit,geller2015tunable,barends2015digital}.

The paper is organized as follows. In Section~\ref{sec:generalized-ssh-model},
we introduce a generalized SSH model, in which each unit cell contains three qubits, and the topology of the model is characterized by winding number. We also use Xmon qubits as an example to show how to form such generalized model. In Section~\ref{sec:2-qubit-entanglment-state}, we derive the edge states of the Hamiltonian given in Section~\ref{sec:generalized-ssh-model}  for the generalized SSH model, and then show that arbitrary two-qubit entangled state can be encoded in these edge states and transferred from one end of the chain to the other through adiabatic process. The exact dynamical solution based on adiabatic theorem is also given. Furthermore, we illustrate the robustness of our proposal against disorder from two parts, i.e., the coupling strength
and evolution time. In Section~\ref{sec:extending-our-formular},
we generalize QST from the entangled state of the two-qubit to those of $\mathcal{N}$-qubit with $\mathcal{N}\geq 3$. In particular, the QST of three-qubit state is carefully analyzed. In Section~\ref{sec:discussions-and-conclution},
we further discuss our proposal and analyze the experimental feasibility, and finally summarize our results.

\section{generalized ssh model with three qubits in each unit cell\label{sec:generalized-ssh-model}}

\begin{figure}[H]
	\begin{centering}
		\includegraphics[width=9cm]{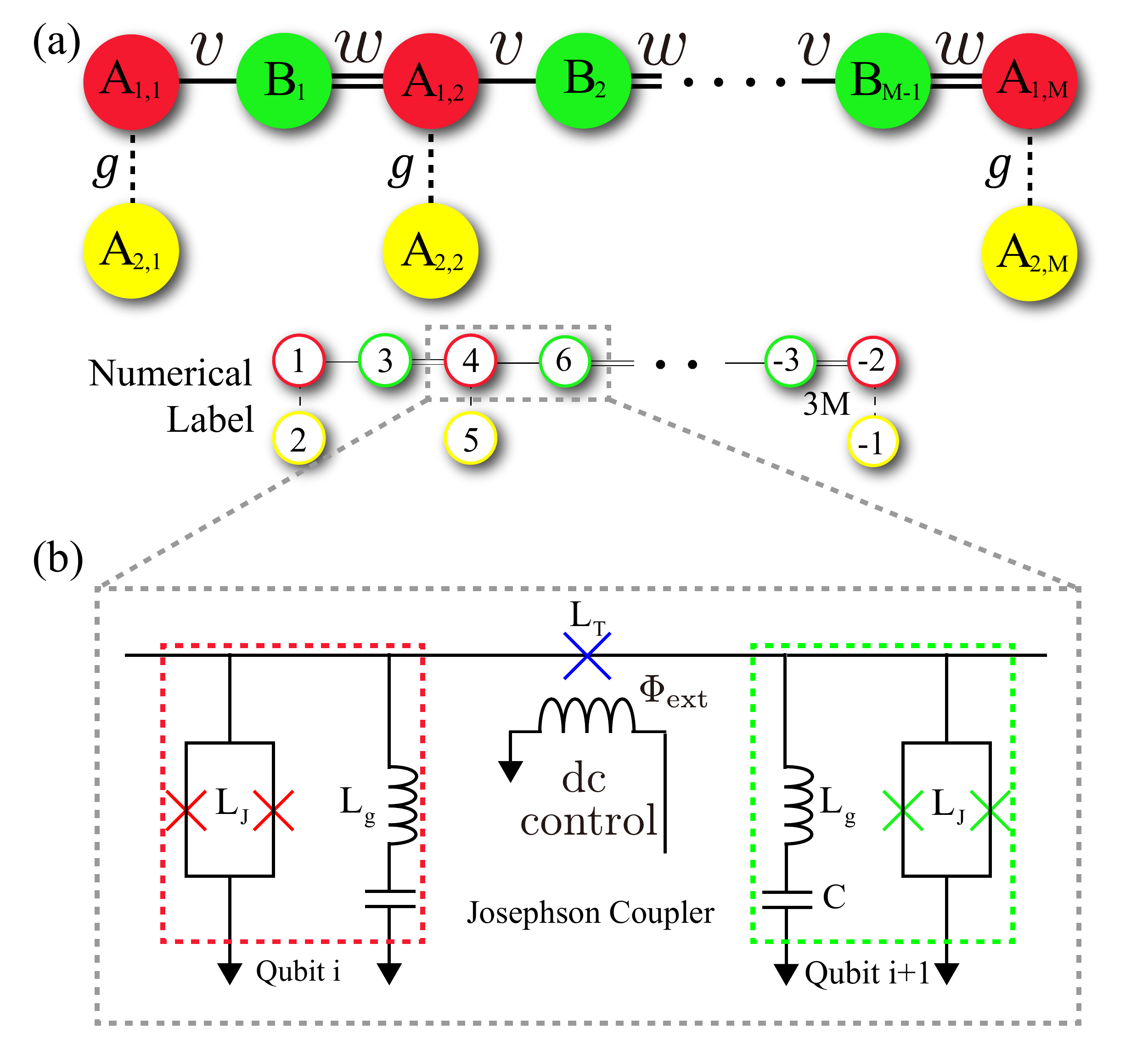}
		\par\end{centering}
	\caption{\label{Fig:Model scheme} (a) Schematic diagram of a one-dimensional qubit chain.
		Each unit cell hosts three qubits, labelled as $A_{1,m}$, $A_{2,m}$ and
		$B_{m}$ separately (The first subscript of $A$ denotes the intracell index. The subscript of $B$ and the second subscript of $A$ denote the intercell index). The coupling strength $g$ between $A_{1,m}$ and $A_{2,m}$ is uniform along the chain, whereas the hopping between $A_{1,m}$ and $B_{m}$ is staggered, denoted as $v$ and $w$.
		Each qubit can also be labeled by one unique numerical index $x$
		in order of $A_{1,1}A_{2,1}B_{1}A_{1,2}\cdots B_{M-1}A_{1,M}A_{2,M}$.
		(b) Realization of the qubit chain with Xmon
		qubits. Each qubit contains three basic superconducting circuits elements,
		i.e, the josephson junction, the capacitance and the inductance. Two
		adjacent qubits are coupled through a Josephson coupler. The qubit-qubit
		coupling can be tuned via an external magnetic flux $\Phi_{\mathrm{ext}}$ with a dc control. The Josephson junctions labeled $\mathrm{L_{J}}$
		are each double junctions threaded by additional fluxes (not shown)
		that tune the qubit frequencies. Therefore, both the qubit frequencies
		and the couplings in the qubit chain are tunable.}
\end{figure}

The SSH model~\citep{su1979solitons} is one of the simplest examples hosting one-dimensional topological phases. This model and its various extensions have been widely used to study different physical phenomena~\citep{perez2018ssh,li2014topological,li2019extended,xu2020general,atala2013direct,li2015winding,meier2016observation,sirker2014boundary}.
In the standard SSH model, each unit cell
has two sublattices. By contrast, for one class of extended SSH models, each unit cell contains 3 or more sublattices, thus called the SSH3 or SSHN models~\citep{xie2019topological,he2020non}. As schematically shown in Fig.~\ref{Fig:Model scheme}(a), we first consider a extended SSH3 model consisting of a qubit chain with $M$ unit cells. Hereafter, we use the unit cell number $M$ to denote the length of the chain.  Each unit cell contains three sublattices labelled as $A_{1,m}$, $A_{2,m}$, and $B_{m}$ with $m=1,2,\cdots, M$. The sublattice $B_{M}$ of the $M$th unit cell is removed from the right end of the chain. Sublattices $A_{1,m}$ and $B_{m}$ are analogous to those in the standard two-band SSH model with staggered coupling strengths $v$ and $w$. An extra sublattice $A_{2,m}$ is coupled to $A_{1,m}$ with the coupling strength $g$ ($g>0$). Thus, the Hamiltonian of such a  qubit chain is ($\hbar=1$)
\begin{eqnarray}
	H & = & \stackrel[m=1]{M-1}{\sum}\left(v\sigma_{A_{1,m}}^{+}\sigma_{B_m}^{-}+w\sigma_{A_{1,m+1}}^{+}\sigma_{B_m}^{-}+\textrm{H.c.}\right)\nonumber \\
	& + & \stackrel[m=1]{M}{\sum}\left(g\sigma_{A_{1,m}}^{+}\sigma_{A_{2,m}}^{-}+\textrm{H.c.}\right).\label{eq:H}
\end{eqnarray}
with $m$ denoting the index of each unit cell. The ladder operators in the $m$th unit cell is given by $\sigma_{I}^{+}=\vert e_{I}\rangle\langle g_{I}\vert$ ($I=A_{1,m},A_{2,m},B_{m}$), with $\vert e_{I}\rangle$ and $\vert g_{I}\rangle$ denoting the excited and ground states, and the operator $\sigma_{I}^{-}$ is the Hermitian conjugate of the operator $\sigma_{I}^{+}$.

We note that our proposal can be applied to any kinds of controllable qubit systems. However, for the concreteness of the studies, we use superconducting qubit circuits, e.g., Xmon qubits, to construct our theoretical model.
This $X$-shape qubit has several advantages, e.g., high-coherence, fast tunable coupling, and easy connection~\citep{chen2014qubit,geller2015tunable,barends2015digital}, thus is more suitable for our proposal.
Our proposal is not limited to the Xmon qubit, and can also be applied to other types of superconducting qubits, e.g., transmon or flux qubit. For our setup, as shown in Fig.~\ref{Fig:Model scheme}(b), two Xmon qubits are connected with each other by a Josephson junction coupler~\citep{chen2014qubit,geller2015tunable}. An extra magnetic flux bias
$\Phi_{\rm ext}$ is applied to tune the effective linear inductance of the coupler junction.
Thus, the coupling constant between these two qubits is tunable with controllable extra magnetic flux $\Phi_{\rm ext}$ (see details in Appendix~\ref{sec:Xmon-qubut-chain}).

In the following studies, we only consider the single-excitation of the chain,
thus the Hamiltonian can be rewritten as
\begin{eqnarray}
H& = & \sum_{m=1}^{M-1}\left(v\vert\mathcal{A}_{1,m}\rangle\langle\mathcal{B}_{m}\vert+w\vert\mathcal{A}_{1,m+1}\rangle\langle\mathcal{B}_{m}\vert+\textrm{H.c.}\right)\nonumber \\
 & + & \sum_{m=1}^{M}\left(g\vert\mathcal{A}_{1,m}\rangle\langle\mathcal{A}_{2,m}\vert+\textrm{H.c.}\right) \label{eq:Hamiltonian-2}
\end{eqnarray}
in the single-excitation subspace \{$\vert\mathcal{A}_{1,m}\rangle,\vert\mathcal{A}_{2,m}\rangle,\vert\mathcal{B}_{m}\rangle$\},
with $\vert\mathcal{A}_{1,m}\rangle=\sigma_{A_{1,m}}^{+}\vert G\rangle$,
$\vert\mathcal{A}_{2,m}\rangle=\sigma_{A_{2,m}}^{+}\vert G\rangle$ and
$\vert\mathcal{B}_{m}\rangle=\sigma_{B_m}^{+}\vert G\rangle$. Here $\vert G\rangle$ denotes that all qubits in the
chain are in the ground state, i.e., $\vert G\rangle=\vert g_{A_{1,1}}g_{A_{2,1}}g_{B_{1}}\cdots g_{A_{2,M}}\rangle$,
which is written as $\vert G\rangle=\vert gg\cdots g\rangle$ for simplicity.

For the standard two-band SSH model, the topologically nontrivial phase is characterized by a nonzero winding number.
Two topological in-gap edge modes are degenerate at zero energy in the thermodynamic limit~\citep{asboth2016short}.
However, for a finite lattice size, the two edge states hybridize due to finite-size effect and so that their energy eigenvalues  are  shifted by  an exponentially small amount.
Dynamics-wise, this edge-state hybridization would then induce
Rabi oscillation between two topological edge modes if the initial
state is prepared by exciting the leftmost or rightmost qubit only~\citep{nie2020bandgap}.
To obtain a steady edge state in a finite-size system, one may remove one edge qubit from the standard SSH chain~\citep{mei2018robust,cai2019observation}. In this case, this imperfect SSH chain only supports one edge state. This is useful for our considerations here since we are considering the same platform.  Indeed, applying this idea of eliminating one edge qubit from an extended SSH model, we can likewise engineer steady edge states in the extended SSH setting, as shown below.

For our extended SSH3 model shown in Fig.~\ref{Fig:Model scheme}(a), the Hilbert space of the Hamiltonian is enlarged by the extra qubits $A_{2,m}$ compared with the standard SSH model. The edge states are expected to form from two renormalized branches of the qubit chain, thus resulting in the upper and the lower edge states with positive and negative eigenenergies, as discussed further in Appendix~\ref{sec:A-straightforward-diagram}.
To see the topological aspect of these edge states, we first consider the case with $g=0$, where
sublattices $A_{1,m}$ and $B_{m}$ are decoupled
from $A_{2,m}$. In this case, the imperfect SSH model with an odd number of qubits is formed, where $v$ and
$w$ are the intra- and inter- cell couplings. In the standard SSH model, edge
states appear when the coupling strength at the edge is weaker than
the coupling next to it. Similarly, for the imperfect SSH model, a weaker coupling strength appears at the left (right) edge and generates an edge state there when $v<w$ ($v>w$). Therefore, in the presence of an odd number of sites, the system always has one edge state in the topologically nontrivial regime~\citep{mei2018robust}. That is, in our extended SSH3 model,  either the left or the right edge state corresponds to a winding number of $1$ for the Fourier transformed Hamiltonian (in momentum space),
with unit cells defined either as sites $(A_{1,m},A_{2,m},B_{m})$ or as sites $(B_{m},A_{1,m+1},A_{2,m+1})$ in Fig.~\ref{Fig:Model scheme}(a) respectively.
When $g=0$, the edge state is only located at the sublattice $A_{1,m}$, which returns to the case of the standard SSH chain. When $g>0$, the edge states are expected to occupy both sublattices $A_{1,m}$ and $A_{2,m}$, because they take the role of the $A$-type sublattice in the standard SSH chain (see Appendix~\ref{sec:Topological-invariant}).
With this understanding, we can see that the edge states in our extended SSH model do originate from the topological edge states in the standard SSH model, and hence should be robust to local disorder.

Below we show how such edge states in the extended SSH chain can be used to robustly transfer arbitrary entangled states. We here emphasize that the parameters used for the following numerical simulations are taken from superconducting qubit circuits, e.g., Xmon qubit circuits~\citep{chen2014qubit,geller2015tunable,barends2015digital}.

\section{2-qubit entanglment state transfer\label{sec:2-qubit-entanglment-state}}

\subsection{Edge states of the qubit chain}

In our extended SSH model, there are $\mathcal{L}=3M-1$ qubits in the chain when
one $B$-type qubit at the right end of the chain is removed. Let us first give an ansatz that the resultant edge state in our system exclusively occupies sublattices $A_{1,m}$ and $A_{2,m}$ (see appendix \ref{sec:A-straightforward-diagram}), i.e., the associated wavefunction can be written as
\begin{equation}
\vert\varPsi_{\rm edge}\rangle=\stackrel[m=1]{M}{\sum}\lambda^{m}\left(a\sigma_{A_{1,m}}^{+}+b\sigma_{A_{2,m}}^{+}\right)\vert G\rangle. \label{eq:psi}
\end{equation}
As energy-eigenstates of the system's Hamiltonian, the wavefunction in Eq.~(\ref{eq:psi}) must satisfy the stationary Schr\"odinger equation
\begin{equation}
H\vert\varPsi_{\rm edge}\rangle=E\vert\varPsi_{\rm edge}\rangle.
\end{equation}
Combining Eq.~\eqref{eq:H} with Eq.~\eqref{eq:psi}, we have
\begin{align}
E\left(a\sigma_{A_{1,m}}^{+}+b\sigma_{A_{2,m}}^{+}\right)\vert G\rangle & =  a\left(v+w\lambda\right)\sigma_{A_{2,m}}^{+}\vert G\rangle  \\
& + g\left(b\sigma_{A_{1,m}}^{+}+a\sigma_{A_{2,m}}^{+}\right)\vert G\rangle. \nonumber
\end{align}
It is straightforward to get $v+w\lambda=0$, i.e., $\lambda=-v/w$,
and the coefficients $a$ and $b$ satisfy the following equation as
\begin{equation}
\left(\begin{array}{cc}
0 & g\\
g & 0
\end{array}\right)\left(\begin{array}{c}
a\\
b
\end{array}\right)=E\left(\begin{array}{c}
a\\
b
\end{array}\right).
\end{equation}
Thus, the energy eigenvalues of the edge states are given by $E_{\pm}=\pm g$, with coefficients ($a,b$) being ($1/\sqrt{2},1/\sqrt{2}$) or ($1/\sqrt{2},-1/\sqrt{2}$) respectively. This explicit edge state solution indicates that there are two edge states with the form of the
Bell state $\vert\chi_{m,\pm}\rangle=\left(\vert\mathcal{A}_{1,m}\rangle\pm\vert\mathcal{A}_{2,m}\rangle\right)/\sqrt{2}$
in each unit cell. The wavefunctions corresponding to the edge states hence take the following form (unnormalized)
\begin{equation}
\vert\varPsi_{\pm}\rangle=\stackrel[m=1]{M}{\sum}\lambda^{m}\left(\frac{\sigma_{A_{1},m}^{+}\pm\sigma_{A_{2},m}^{+}}{\sqrt{2}}\right)\vert G\rangle,\label{eq:psi-1}
\end{equation}
where both the upper (labeled as $+$) and lower (labeled as $-$)
edge states only occupy the $A$-type qubits. When $\left|\lambda\right|=v/w\ll1$, i.e., $v\ll w$, the edge states are mostly localized in the left end of the chain as
\begin{equation}
\vert\varPsi_{\pm}\rangle\approx\vert L_{\pm}\rangle=\frac{1}{\sqrt{2}}\left(\sigma_{A_{1,1}}^{+}\pm\sigma_{A_{2,1}}^{+}\right)\vert G\rangle.
\end{equation}
In the limit of $v=0$, the qubit chain degenerates into $M-1$  trimers and an additional dimer,  an exact edge state $\vert L_{\pm}\rangle$ can be obtained at the left end of the chain.
When $\left|\lambda\right|=v/w\gg1$, i.e., $v\gg w$, the edge states
are mostly localized in the right end of the chain as
\begin{equation}
\vert\varPsi_{\pm}\rangle\approx\vert R_{\pm}\rangle=\frac{1}{\sqrt{2}}\left(\sigma_{A_{1,M}}^{+}\pm\sigma_{A_{2,M}}^{+}\right)\vert G\rangle.
\end{equation}
In the limit of $w=0$, an exact edge state $\vert R_{\pm}\rangle$ can be obtained at the right end of the chain.

In Fig.~\ref{Fig:two-qubit state}(a), as an example, the energy
spectrum of a $14$-qubit chain with $5$ unit cells is
plotted for $\left|\lambda\right|=v/w\in\left[0,\infty\right]$. It clearly shows that the two topological edge modes (color-dashed lines) exist in the gaps of three bulk bands. The bulk band in the middle of two edge modes is 4-fold degenerate as shown in Fig~\ref{Fig:two-qubit state}(b). As the parameter
$\lambda$ changes from $0$ to $\infty$, the localized topological edge states as a function of $\lambda$ also change their qualitatively behavior, namely,
from being localized at the left end (red) to being localized at the right end (green). Meanwhile, the eigenvalues $E_{\pm}$ corresponding to these two edge states keep constants all along.

\subsection{QST process for arbitrary two-qubit entangled states}

\begin{figure}[htp]
	\begin{centering}
		\includegraphics[width=8.5cm]{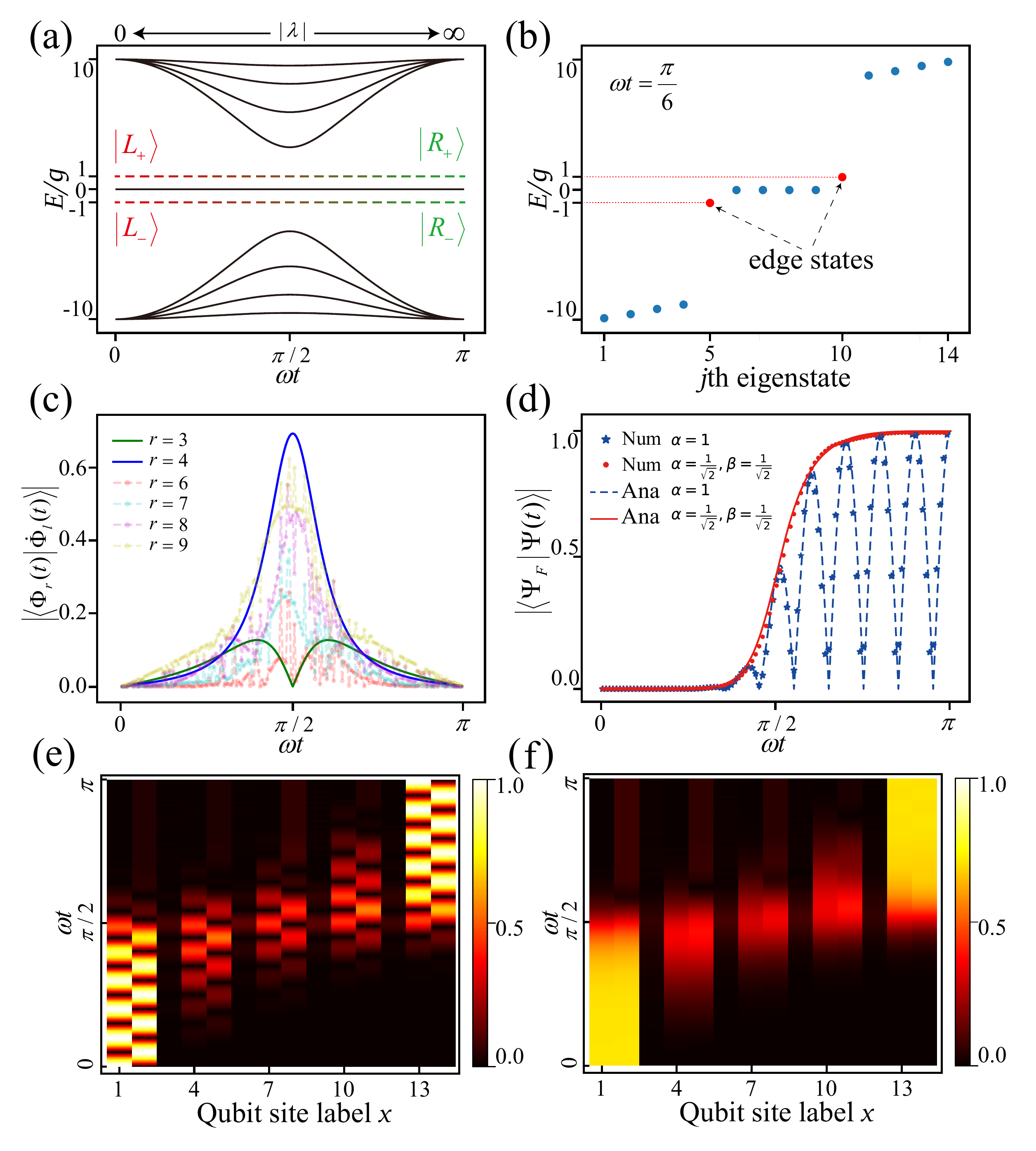}
		\par\end{centering}
	\caption{\label{Fig:two-qubit state}Two-qubit state transfer via a 14 qubits chain with $5$ unit cells, i.e., $M=5$. The system parameters are $J=5g$ and $g/2\pi=10$MHz. (a) The energy spectrum for an extended SSH3 model. Black-solid
		lines represent the bulk states and color-dashed lines represent the
		edge states. Three bulk bands are divided by two edge states, the middle bulk is $4$-fold degenerate.
		As $\omega t$ changes from $0$ to $\pi$, edge states are transferred from the
		left end (red) to the right end (green). (b) Schematics for eigenstates distribution at $\omega t=\pi/6$. Each point represents one eigenstate. Blue and red dots represent bulk and edge states, respectively. (c) The variation of $|\langle \Phi_{r}(t)|\dot{{\Phi}}_{l}(t)\rangle|$ with the time evolution. Here $\Phi_{l}(t)$ takes the lower edge state, i.e., $l=5$.  (d) Time evolutions of the target state occupations $\left|\langle\varPsi_{F}\vert\varPsi\left(t\right)\rangle\right|$ for different initial states.  Red-solid and blue-dashed curves denote theoretical solutions.
		Red dots and blue stars denote the numerical simulations. The red-solid curve and red dots correspond to
		the initial state $\left(\vert \mathcal{A}_{1,1}\rangle+\vert \mathcal{A}_{2,1}\rangle\right)/\sqrt{2}$. Meanwhile,
		the blue-dashed curve and blue stars correspond to the initial state $\vert \mathcal{A}_{1,1}\rangle$. (e) Time-dependent population distribution on each qubit in the chain when the initial state is prepared  to  single-qubit state $\vert \mathcal{A}_{1,1}\rangle$.  (f) Time-dependent population distribution on each qubit in the chain when the initial state is prepared to two-qubit Bell state $\left(\vert \mathcal{A}_{1,1}\rangle+\vert \mathcal{B}_{2,1}\rangle\right)/\sqrt{2}$.}
\end{figure}

According to the quantum adiabatic theorem, if the parameters
of the qubit chain in Eq.~\eqref{eq:H} can be changed slowly enough, it will remain in its instantaneous
eigenstate. At $v=0$, the leftmost two qubits $A_{1,1}$ and $A_{2,1}$ are isolated
and the edge states are $\vert L_{\pm}\rangle=\vert\chi_{1,\pm}\rangle\vert gg\cdots g\rangle$.
At $w=0$, the rightmost two qubits $A_{1,M}$ and $A_{2,M}$ are isolated and the edge states
are $\vert R_{\pm}\rangle=\vert gg\cdots g\rangle\vert\chi_{M,\pm}\rangle$. It is obvious that the coupling strengths
\begin{equation}
	v=J\left[1-\cos\left(\omega t\right)\right],w=J\left[1+\cos\left(\omega t\right)\right],\label{eq:vw}
\end{equation}
can be changed slowly, then the state of the left end can be transferred to the right end of the chain.
Where $\omega$ and $J$ are the frequency and strength of the control field, respectively.
For example, as shown in Fig.~\ref{Fig:two-qubit state}(a), if the system is initially prepared to the state $\vert L_{\pm}\rangle$ for the time $t=0$, then this state adiabatically evolves from $\vert L_{\pm}\rangle$ to $\vert R_{\pm}\rangle$ when the time $t$ slowly varies from $0$ to $\pi/\omega$. The adiabatic following is feasible here because the considered state is the system's edge state, which is gapped from the bulk states.

In above, we analyze QST when the edge state is the initial state. Let us now consider the general situation that the system is initially prepared to an arbitrary entangled state
\begin{equation}
	\vert\varPsi_{\rm in}\rangle=\left(\alpha\sigma_{A_{1,1}}^{+}+\beta\sigma_{A_{2,1}}^{+}\right)\vert G\rangle,\label{eq:psi-1-1}
\end{equation}
 at the left end of the chain,  which can be decomposed as
\begin{equation}
	\vert\varPsi_{\rm in}\rangle=\frac{\alpha+\beta}{\sqrt{2}}\vert L_{+}\rangle+\frac{\alpha-\beta}{\sqrt{2}}\vert L_{-}\rangle.\label{eq:10}
\end{equation}
For such more general situation involving the superpositions of two edge states, one must carefully analyze the quantum phases. In an adiabatic process, if the system is initially prepared to an eigenstate $\vert\varPhi\left(0\right)\rangle$, then the final state at the time $t$ is given as
\begin{equation}
\vert\psi_{\rm ad}\left(t\right)\rangle=e^{ir\left(t\right)}e^{i\theta\left(t\right)}\vert\varPhi\left(t\right)\rangle,
\end{equation}
where $\theta\left(t\right)=-\int_{0}^{t}E(t^{\prime})dt^{\prime}$ is the dynamical phase
and $r\left(t\right)=i\int_{0}^{t}\langle\varPhi\left(t^{\prime}\right)\vert\dot{\varPhi}\left(t^{\prime}\right)\rangle dt^{\prime}$
is the geometric phase, which can be gauged out unless the evolution
path is closed. The normalization of $\vert\varPhi\left(t^{\prime}\right)\rangle$
implies that $\langle\varPhi\left(t^{\prime}\right)\vert\dot{\varPhi}\left(t^{\prime}\right)\rangle$
is imaginary, which guarantees that $r\left(t\right)$ is real~\citep{berry1984quantal}.
In our protocol, with the chosen gauge as shown in Eq.~\eqref{eq:psi-1},
$\vert\varPsi_{\pm}\rangle$ only contains real parameters along $\lambda\left(t\right)=\left[1-\cos\left(\omega t\right)\right]/\left[1+\cos\left(\omega t\right)\right]$,
thus $\langle\varPhi\left(t^{\prime}\right)\vert\dot{\varPhi}\left(t^{\prime}\right)\rangle$
is real, which guarantees that $r\left(t\right)$ is imaginary.
Therefore, the geometric phase $r\left(t\right)$ must be zero and is naturally gauged
out.

We thus only need to consider the dynamical phase associated with the adiabatic process. If we adiabatically change the parameters $v$ and $w$ to the final time $t$, following the quantum adiabatic theorem, the initial state in Eq.~(\ref{eq:10}) should evolve to the state
\begin{eqnarray}
\vert\varPsi\left(t\right)\rangle&=&\frac{\alpha+\beta}{\sqrt{2}}\vert\varPsi_{+}(t)\rangle e^{-i\int_{0}^{t}E_{+}dt'}\nonumber\\
 & +&
\frac{\alpha-\beta}{\sqrt{2}}\vert\varPsi_{-}(t)\rangle e^{-i\int_{0}^{t}E_{-}dt'}. \label{eq:Analytical solution}
\end{eqnarray}
As discussed above,  $E_{\pm}$ in our model are both constants due to the nature of the topological edge state. As such the final state at the time $t$ is
\begin{equation}
\vert\varPsi\left(t\right)\rangle=\frac{\alpha+\beta}{\sqrt{2}}\vert\varPsi_{+}(t)\rangle e^{-igt}+\frac{\alpha-\beta}{\sqrt{2}}\vert\varPsi_{-}(t)\rangle e^{igt}.\label{eq:analy}
\end{equation}
Indeed, by solely observing the values of $E_{\pm}=\pm g$, we find that the phase factors $e^{-igt}$  and $e^{igt}$ for two involved edge states (from upper and lower branch respectively) have the same period $T=2\pi/g$.
Therefore, if the time evolution takes the dynamical period $T$, the dynamical phase difference between the two involved edge states in Eq.~(\ref{eq:analy}) will be zero. We can call this dynamical period
as one evolution circle. If the total adiabatic protocol time is always chosen to be a multiple of evolution circles, then the concern with dynamical phases can be lifted.

As shown in Fig.~\ref{Fig:two-qubit state}(a), if the control field applied to the coupling strength involves from $t_{0}=0$ to $t_{f}=\pi/\omega$, then the edge state is transferred from the left end to the right end. In this case, Eq.~(\ref{eq:analy}) becomes
\begin{equation}
\vert\varPsi_{f}\rangle=\frac{\alpha+\beta}{\sqrt{2}}\vert R_{+}\rangle e^{-i\pi \frac{g}{\omega}}+\frac{\alpha-\beta}{\sqrt{2}}\vert R_{-}\rangle e^{i\pi \frac{g}{\omega}}.\label{eq:Finalstate-1}
\end{equation}
Here we choose the evolution time $t_{f}$ to be a multiple of evolution circles, i.e., $t_{f}/T=g/2\omega=n$. $n$ should be a large integer number to satisfy the adiabatic condition. Thus, the dynamical phases become zero and the state in Eq.~(\ref{eq:Finalstate-1}) is involved to
\begin{equation}
\vert\varPsi_{F}\rangle=\left(\alpha\sigma_{A_{1,M}}^{+}+\beta\sigma_{A_{2,M}}^{+}\right)\vert G\rangle. \label{eq:14}
\end{equation}
Here, the subscript $F$ denotes the state transferred perfectly, i.e., the final state. Therefore, an arbitrary two-qubit entangled state can be encoded by two edge states and perfectly transferred from the left end to the right end via an adiabatic passage. Notice that  all along the protocol, only the $A_{1,m}$- and $A_{2,m}$-type qubits are occupied by the edge states, and the $B_{m}$-type qubits serve as the invariable medium. Thus the qubits $A_{1,m}$ and $A_{2,m}$ can be considered as the transport qubits, and the qubits $B_{m}$ can be considered as the mediated qubits.

We here have two remarks. First, for the special case of two-qubit state transfer, the evolution
time can be half of the evolution cycle defined above. In this case, the final state acquires a global phase factor $-1$, which does not affect the task of entangled state transfer. Second, one may worry about the robustness of such adiabatic protocol as we need a precise timing.  This is unnecessary because along the way, the eigenvalues of the edge states are pinned at special constant values due to topological features and hence we still expect to have robust protocols.

\subsection{Analysis on adiabatic condition and two examples of QST}

We now analyze the condition of the adiabatic evolutions. The adiabatic approximation requires a small changing rate of the Hamiltonian $\dot{H}\left(t\right)$ and a large energy gap $\left|E_{r}-E_{l}\right|$ between the $r$th and $l$th eigenstates. For our protocol, if we assume that the $l$th eigenstate $\vert\varPhi_{l}\left(t\right)\rangle$ is the edge state, then the adiabatic condition is given by
	\begin{equation}
		\left|\langle\varPhi_{r}\left(t\right)\vert\dot{\varPhi}_{l}\left(t\right)\rangle\right|=\left|\frac{\langle\varPhi_{r}\left(t\right)\vert\dot{H}\left(t\right)\vert\varPhi_{l}\left(t\right)\rangle}{E_{r}\left(t\right)-E_{l}\left(t\right)}\right|\ll1, \label{eq:adiabatic condition}
	\end{equation}
where $E_{r}\left(t\right)$ is the instantaneous eigenenergy corresponding to the instantaneous state $\vert\varPhi_{r}\left(t\right)\rangle$ for the time $t$. The eigenstates are sorted according to the corresponding eigenenergies (lowest to highest), and the $5$th eigenstate in Fig.~\ref{Fig:two-qubit state}(a) is the lower edge state. Here we make the 14-qubit chain evolve 5 evolution circles, i.e., $\omega=0.1g$. As shown in Fig.~\ref{Fig:two-qubit state}(c), for the lower edge state ($l=5$), the adiabatic conditions are checked to be satisfied with $r=3,4,6,7,8,9$. For the bulk state between two edge states ($r=6,7,8,9$), the results are not continuous due to the numerical instablity of degenerate eigensolutions of the system.

To interpret our protocol, we choose the final state occupation $\left|\langle\varPsi_{F}\vert\varPsi\left(t\right)\rangle\right|$ as the dynamical indicator. In Fig.~\ref{Fig:two-qubit state}(d), we compare the analytical and numerical results of this indicator. The analytical solution for $|\Psi(t)\rangle$ is given in Eq.~(\ref{eq:analy}), while the numerical solution is calculated with ordinary differential equation (ODE) solver for small and fixed step size. Here, $\omega=0.1g$, and the evolving time is $t_{f}=0.5$$\mu$s, i.e., exact $5$ times of
the dynamical period ($T=2\pi/g=0.1$$\mu s$). We typically choose two initial states to be transferred. The first one is single-qubit state
\begin{equation}
	\vert\varPsi_{\rm in}^{\left(1\right)}\rangle=\sigma_{A_{1,1}}^{+}\vert G\rangle \equiv \frac{\sqrt{2}}{2}\left(\vert L_{+}\rangle+\vert L_{-}\rangle\right)
\end{equation}
which can be obtained from Eq.~(\ref{eq:10}) by setting $\alpha=1$ and $\beta=0$.
The other one is two-qubit Bell state
\begin{equation}
	\vert\varPsi_{\rm in}^{\left(2\right)}\rangle=\frac{\sqrt{2}}{2}\left(\sigma_{A_{1,1}}^{+}+\sigma_{A_{2,1}}^{+}\right)\vert G\rangle \equiv\vert L_{+}\rangle
\end{equation}
with $\alpha=\beta=1/\sqrt{2}$. For the first one of single-qubit state transfer, the analytical evolution of the state is given as
\begin{equation}
	\vert\Psi^{\left(1\right)}(t)\rangle=\frac{\sqrt{2}}{2}\left(\vert \varPsi_{+}\rangle e^{-igt}+\vert\varPsi_{-}\rangle e^{igt}\right)\label{eq:first-case}
\end{equation}
The analytical solution (blue-dashed line) and numerical simulation (blue stars) for the dynamical indicator $\left|\langle\varPsi_{F}\vert\varPsi\left(t\right)\rangle\right|$ are plotted as a function of the evolution time $t$ in Fig.~\ref{Fig:two-qubit state}(d). In this case, the indicator $\left|\langle\varPsi_{F}\vert\varPsi\left(t\right)\rangle\right|$ has a periodic variation due to the dynamical phase difference between the two edge states. The analytical result agrees well with numerical one, this confirms that the chosen parameters have fulfilled the adiabatic conditions. For two-qubit Bell state
transfer, the adiabatic evolution of the state is
\begin{equation}
	\vert\varPsi^{\left(2\right)}\left(t\right)\rangle=\vert \varPsi_{+}\rangle e^{-igt}.\label{eq:second-case}
\end{equation}
In this case, the dynamical phase can be gauged out as
one global phase, so the dynamical indicator does not oscillate and increases smoothly with the time. Again, the analytical result (red line) agrees well with numerical (red dots) one.

The first case for single-qubit state transfer of Eq.~(\ref{eq:first-case}) is further
analyzed in Fig.~\ref{Fig:two-qubit state}(e), where
we show the population distribution $|\langle P_{m}|\Psi^{\left(1\right)}(t)\rangle|$ ($P_{m} \in \{\vert\mathcal{A}_{1,m}\rangle,\vert\mathcal{A}_{2,m}\rangle,\vert\mathcal{B}_{m}\rangle\}$) on each site of the state $|\Psi^{\left(1\right)}(t)\rangle$ during the adiabatic protocol.
The state distribution shifts from the left to the right with rapid oscillations, due to coherent effect caused by the dynamical phase difference between the two involved edge states as mentioned above. However, for the second case shown in Eq.~(\ref{eq:second-case}), Fig.~\ref{Fig:two-qubit state}(f) demonstrates the smooth transfer of the state $|\Psi^{\left(2\right)}(t)\rangle$ without any oscillation.

\subsection{Robustness analysis for two-qubit state transfer}

Any realistic implementation of the theoretical protocol unavoidably involves disorders from many aspects, e.g., the environmental effect, the time inaccuracy of the control field applied for the adiabatic evolution, nonuniform of the prepared qubits and imprecise couplings between the qubits in the chain. Here, we consider two main imperfections: one is the disorder of the qubit couplings and the other one is inaccuracy of the evolution time to achieve perfect state transfer.

The first kind of the disorders can be analyzed by modelling it as external
perturbation terms in our system Hamiltonian, i.e.,
\begin{align}
\delta H & =  \stackrel[m=1]{M-1}{\sum}\left(\delta\mu_{A_{1,m}}\sigma_{A_{1,m}}^{+}\sigma_{B_m}^{-}+\delta\mu_{B_m}\sigma_{A_{1,m+1}}^{+}\sigma_{B_m}^{-}+\textrm{H.c.}\right)\nonumber \\
 & +  \stackrel[m=1]{M}{\sum}\left(\delta\mu_{A_{2,m}}\sigma_{A_{1,m}}^{+}\sigma_{A_{2,m}}^{-}+\textrm{H.c.}\right),\label{eq:H-1}
\end{align}
where $\delta\mu_{A_{1,m}}$, $\delta\mu_{A_{2,m}}$, and $\delta\mu_{B_{m}}$ are the disorder coefficients, and assumed to satisfy the the Gaussian distribution as $\sim \exp[(-\delta\mu)^2/2\xi^2]$, with
$\xi$ being the standard deviation of the disorder in the coupling strength, i.e., the coupling disorder strength. Note that all these disorders are related to the control field. Thus, with the variation of the adiabatic
parameters, it is more realistic to assume these disorders to be time-dependent, i.e., the temporal
noises. If the initial state is prepared to $\vert\varPsi_{\rm in}\rangle=\left(\alpha\sigma_{A_{1},1}^{+}+\beta\sigma_{A_{2},1}^{+}\right)\vert G\rangle$,
then the adiabatic evolution can be derived as
\begin{equation}
\vert\psi\left(t\right)\rangle=U\left(t\right)\vert\varPsi_{\rm in}\rangle=\mathcal{T}e^{-i\int_{0}^{t}H\left(t'\right)+\delta H\left(t'\right)dt'}\vert\varPsi_{\rm in}\rangle, \label{eq:16}
\end{equation}
where $\mathcal{T}$ is the time order operator.  The numerical
simulation of this process can be written as $U\left(t\right)=\mathcal{T}\prod e^{-i\left[H\left(t'\right)+\delta H\left(t'\right)\right]\Delta t}$ ($\Delta t\ll T$). When $t=t_{f}$, the state described in Eq.~(\ref{eq:16}) evolves to $\vert\psi_{f}\rangle$.  Thus, the fidelity is given as
\begin{equation}
F=\left|\langle\varPsi_{F}\vert\psi_{f}\rangle\right|,\label{eq:17}
\end{equation}
where $\vert\varPsi_{F}\rangle$ is the perfectly transferred state given in Eq.~(\ref{eq:14}).

\begin{figure}[htb]
	\begin{centering}
		\includegraphics[width=9cm]{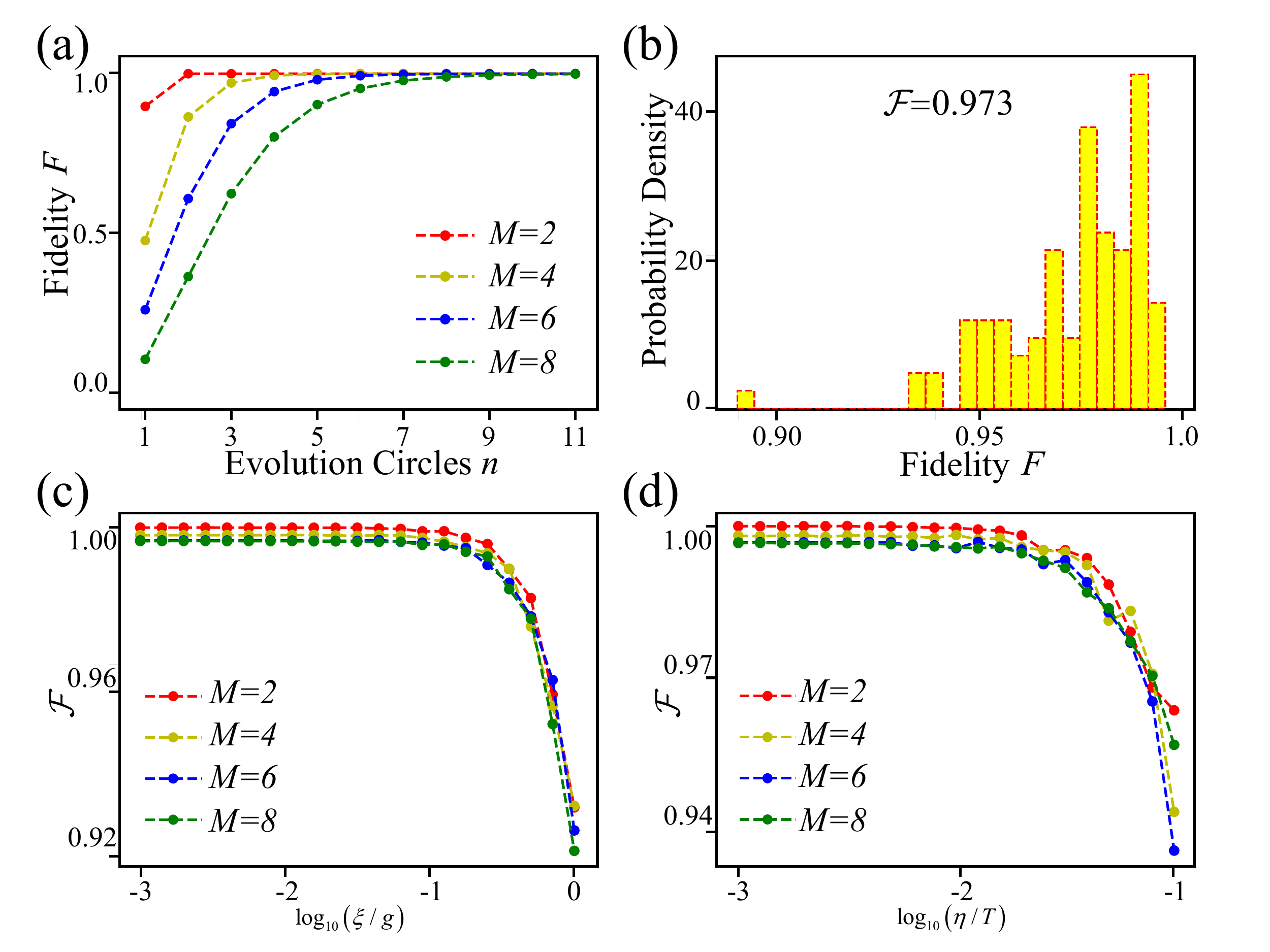}
		\par\end{centering}
	\caption{\label{Fig:two-qubit disorder} Robustness of our protocol for two-qubit
		state transfer. (a) The right edge target-state occupations (fidelity $F$) with different
		evolving circles. As the qubit number of the chain becomes larger,
		the evolving circles demanded for the adiabatic evolution increase
		from 5 to 10. For a qubit chain containing  23 qubits, i.e., 8 unit cells, it needs at least 10 evolving
		circles to guarantee the adiabaticity. (b) The distribution of fidelities
		for one settled state transfer process with 100 repetitions where
		$\xi=0.5g$ and $M=4$. (c)
		The average fidelities of two-qubit entangled state transfer with
		the coupling disorder. The numbers of unit cells are 2,  4, 6, and 8 for
		each curve, separately. (d) The average fidelities of two-qubit entangled
		state transfer with the imperfection of the evolution time.}
\end{figure}

To verify the protocol robustness against these disorders, we first determine the proper evolution circles $n$ under the ideal condition without disorder. As shown in Fig.~\ref{Fig:two-qubit disorder}(a),
we have numerically calculated the fidelity $F$ given in Eq.~(\ref{eq:17}) for the right edge state transferred  with different evolution circles and different lengths of the chain. These results help us to
determine what extent we are working in the adiabatic regime. Indeed, the fidelity rises rapidly with the increases in the number of evolution circles. As shown in Fig.~\ref{Fig:two-qubit disorder}(a), when the number of unit cells of the chain varies from $2$ to $8$, the necessary numbers of evolution circles to achieve good fidelity increase from $5$ to $10$. Therefore, to
guarantee the adiabatic condition and obtain good fidelity for the transferred state, we choose $n=10$ as the number of evolution
circles for the following discussions, i.e., $t_{f}=10\,T=1\mu s$.

In our numerical simulations, we do $100$ repetitions of the adiabatic evolution for a given time $t_{f}$, and each repetition has its different random choice of disorder. We here consider the average fidelity $\mathcal{F}=\overline{F}$ over these $100$ calculations. In Fig.~\ref{Fig:two-qubit disorder}(b), we plot the distribution of fidelities for $100$ repetitions of one chosen state transfer protocol. Here, the number of unit cells is taken as $M=4$, and the coupling disorder strength $\xi$ is taken as $\xi=0.5g$. The distribution of fidelities for each simulation run
is highly concentrated around the average fidelity, so the average fidelity here
is a proper representation for the protocol fidelity.

In Fig~\ref{Fig:two-qubit disorder}(c), the effect of the coupling disorder on the protocol fidelities is shown for different lengths of the chain, i.e., $M=2,4,6,8$, respectively.  For each length of the chain, we find that the average fidelity $\mathcal{F}$ for $100$ repetition calculations decreases from $1$ to $0.92$,
as the coupling disorder strength $\xi$ changes from $10^{-3}g$ to $g$. We also find that there is a plateau near $\mathcal{F}=1$ for each case when $\xi$ is taken from the range $\xi\in\left[10^{-3}g,10^{-1}g\right]$. Therefore, for our protocol, the fidelity $\mathcal{F}$ will remain above $99\%$ as long as the coupling disorder strength $\xi$ is less than $0.1g$, which should be experimentally accessible.

The second kind of the disorders is the inaccuracy of protocol execution time. Unlike a usual adiabatic process, our protocol demands the
total evolution time be exact multiple of the dynamical period. However,
the control inaccuracy for the evolution time may lead to the imperfection of the target
state. This kind of imperfection can be examined by an external perturbation time $\delta t$, which satisfies the Gaussian distribution as $\delta t\sim \exp[(-\delta t)^2/2\eta^2]$.
$\eta$ is the standard deviation of the evolution time. In this case, the modified transferred
state can be derived as
\begin{equation}
\vert\psi_{f}\rangle=U\left(t_{f}+\delta t\right)\vert\varPsi_{\rm in}\rangle=\mathcal{T}e^{-i\int_{0}^{t_{f}+\delta t}H\left(t'\right)dt'}\vert\varPsi_{\rm in}\rangle,
\end{equation}
with the time inaccuracy $\delta t$. Figure~\ref{Fig:two-qubit disorder}(d) depicts
the impact of the evolution time imperfection. We find that the average
fidelity $\mathcal{F}$ for each chain decreases from $1$ to $0.94$ as the time disorder
strength $\eta$ changes from $10^{-3}T$ to $10^{-1}T$. For our protocol,
the fidelity will remain above $99\%$ when the time disorder strength
$\eta$ is less than $0.01T$.

Therefore, we conclude that our protocol is robust against both the qubit coupling disorder and the inaccuracy of the protocol execution time.

\section{Extended protocols for $\mathcal{N}$-qubit state transfer\label{sec:extending-our-formular}}


\begin{figure}[H]
\begin{centering}
\includegraphics[width=9cm]{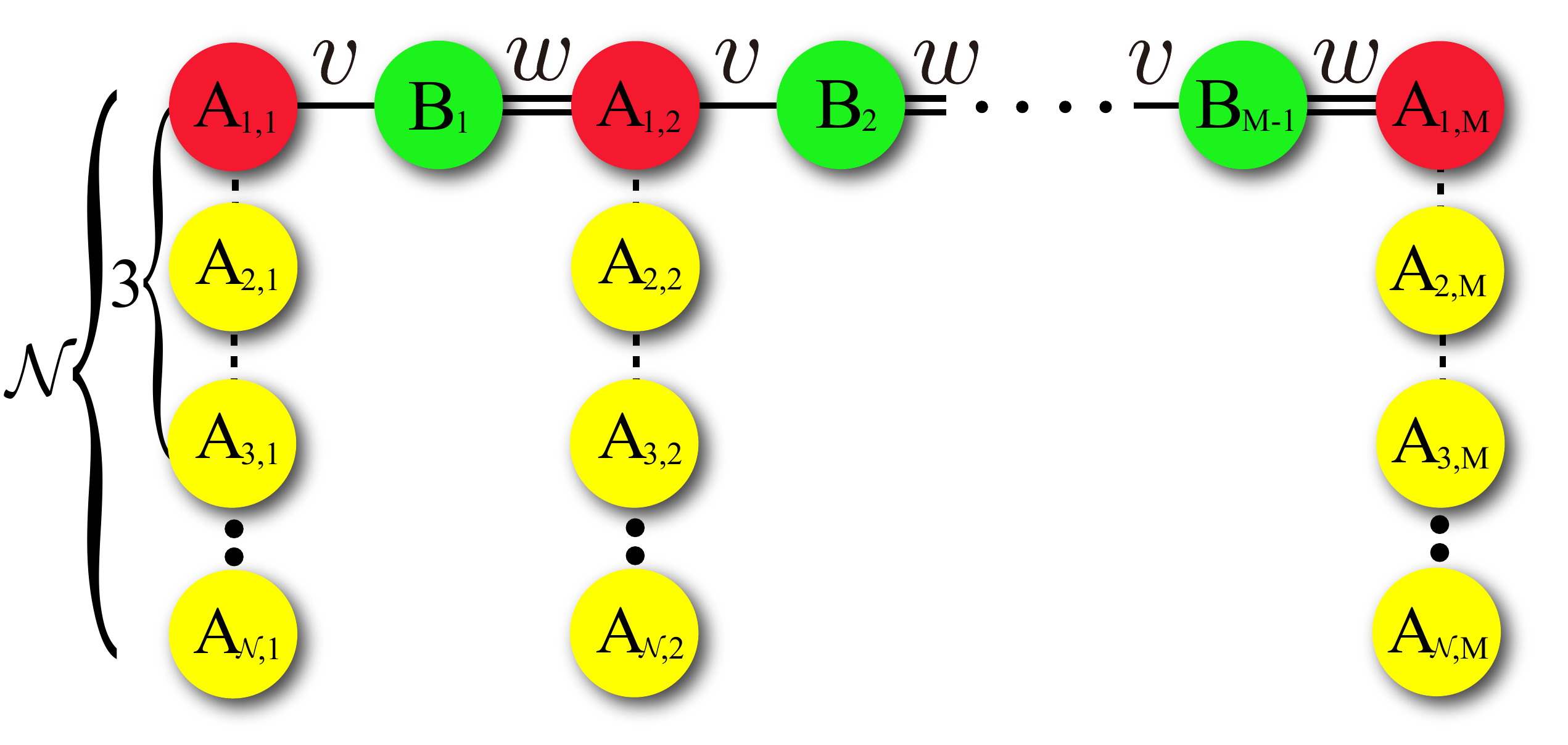}
\par\end{centering}
\caption{\label{Fig:N-qubit state transfer}Schematic of the extended SSH chain
for $3$-qubit and $\mathcal{N}$-qubit state transfer. Each unit cell
hosts $4$ to $\mathcal{N}+1$ qubits. The edge states are localized
among the transport qubits from $A_{1,m}$ to $A_{\mathcal{N},m}$ with $m=1,\cdots, M$. The rest qubits are mediated qubits labeled as $B_{m}$ with $m=1,\cdots, M-1$. $v$ denotes the coupling between
qubits $A_{1,m}$ and $B_{m}$, $w$ denotes the coupling between qubits $A_{1,m+1}$ and $B_{m}$.}
\end{figure}

Our protocol for arbitrary two-qubit  entangled state transfer through extended SSH chain can be easily generalized to  $\mathcal{N}$-qubit state transfer. For $\mathcal{N}$-qubit transfer process, as
schematically shown in Fig.~\ref{Fig:N-qubit state transfer}, the extended SSH model has $\mathcal{N}+1$ sites ($\mathcal{N}$ transport qubits and $1$ mediated qubit) in each unit cell, and the whole qubit chain with $M$ unit cells has $\mathcal{L}=\left(\mathcal{N}+1\right)M-1$ qubits. With the increase of the site number in each unit cell, more edge states emerge. Thus it is not an easy task to find a proper evolution time to cancel all dynamical phase differences between these edge states in our protocol.  As such, the couplings between the transport qubits should not simply be the constant $g$.  To this end,
we may consider the following modified  Hamiltonian
\begin{eqnarray}
H & = &  \stackrel[m=1]{M-1}{\sum}\left(v\sigma_{A_{1,m}}^{+}\sigma_{B_m}^{-}+w\sigma_{A_{1,m+1}}^{+}\sigma_{B_m}^{-}+\textrm{H.c.}\right)\label{eq:H-N}\nonumber\\
& + &  \stackrel[m=1]{M}{\sum}\left(g_{1}\sigma_{A_{1,m}}^{+}\sigma_{A_{2,m}}^{-}+g_{2}\sigma_{A_{2,m}}^{+}\sigma_{A_{3,m}}^{-}+\cdots\right.\nonumber\\
& + & \left. g_{\mathcal{N}-1}\sigma_{A_{\mathcal{N}-1,m}}^{+}\sigma_{A_{\mathcal{N},m}}^{-}+\textrm{H.c.}\right).
\end{eqnarray}
In each unit cell, these $\mathcal{N}$ transport quibts form an array with coupling constants $g_{1}, \cdots, g_{\mathcal{N}-1}$, and all of the mediated qubits $B_{m}$ are used to couple the arrays of transport qubits with coupling constants $v$ and $w$, respectively.

\subsection{$3$-qubit state transfer}

Let us now further illustrate our proposal for $3$-qubit entangled state transfer, i.e., $\mathcal{N}=3$ as an example. In this case, the qubit chain contains $\mathcal{L}=4M-1$ qubits and the corresponding Hamiltonian is given by Eq.~(\ref{eq:H-N}) with $\mathcal{N}=3$ as
\begin{eqnarray}
	H & = & \stackrel[m=1]{M-1}{\sum}\left(v\sigma_{A_{1,m}}^{+}\sigma_{B_m}^{-}+w\sigma_{A_{1,m+1}}^{+}\sigma_{B_m}^{-}+\textrm{H.c.}\right)\nonumber \\
	& + & \stackrel[m=1]{M}{\sum}\left(g\sigma_{A_{1,m}}^{+}\sigma_{A_{2,m}}^{-}+g\sigma_{A_{2,m}}^{+}\sigma_{A_{3,m}}^{-}+\textrm{H.c.}\right).\label{eq:3-H}
\end{eqnarray}
The edge states are only localized in all $A$-type qubits (i.e., $A_{1,m}$-, $A_{2,m}$-, and $A_{3,m}$-type) and can be expanded as
\begin{equation}
\vert\varPsi_{\rm edge}\rangle=\stackrel[m=1]{M}{\sum}\lambda^{m}\left(a\sigma_{A_{1,m}}^{+}+b\sigma_{A_{2,m}}^{+}+c\sigma_{A_{3,m}}^{+}\right)\vert G\rangle.\label{eq:3-edgestates}
\end{equation}
Substituting this equation into eigen-energy function $H\vert\varPsi_{\rm edge}\rangle=E\vert\varPsi_{\rm edge}\rangle$,
we have
\begin{align}
E & \left(a\sigma_{A_{1,m}}^{+}+b\sigma_{A_{2,m}}^{+}+c\sigma_{A_{3,m}}^{+}\right)\vert G\rangle\nonumber \\
& = g\left(b\sigma_{A_{1,m}}^{+}+a\sigma_{A_{2,m}}^{+}+c\sigma_{A_{2,m}}^{+}+b\sigma_{A_{3,m}}^{+}\right)\vert G\rangle\nonumber \\
& + a\left(v\sigma_{B_m}^{+}+w\lambda\sigma_{B_m}^{+}\right)\vert G\rangle.
\end{align}
Here, for this special case of 3-qubit entangled state transfer, we have assumed that the coupling constant $g_{1}$ between $A_{1,m}$-qubit and $A_{2,m}$-qubit equals to that $g_{2}$  between $A_{2,m}$-qubit and $A_{3,m}$-qubit, i.e., $g_{1}=g_{2}=g$.

It is straightforward to obtain $\lambda=-v/w$, and the coefficients
$a$, $b$, and $c$ satisfy the following eigen-equation
\begin{equation}\label{eq:25}
\left(\begin{array}{ccc}
0 & g & 0\\
g & 0 & g\\
0 & g & 0
\end{array}\right)\left(\begin{array}{c}
a\\
b\\
c
\end{array}\right)=E\left(\begin{array}{c}
a\\
b\\
c
\end{array}\right).
\end{equation}
Solving Eq.~(\ref{eq:25}), we can obtain three eigenvalues $E_{\pm}=\pm\sqrt{2}g$ and $E_{0}=0$, corresponding
to three eigenstates $\left(1/2,\pm\sqrt{2}/2,1/2\right)$ and $\left(1/\sqrt{2},0,-1/\sqrt{2}\right)$, respectively.  These three eigenvalues are also eigenenergies of three edge states.  Thus,
three edge states are constructed by the eigenstates, i.e.,  $\vert\chi_{m,\pm}\rangle=\left(\vert\mathcal{A}_{1,m}\rangle\pm\sqrt{2}\vert\mathcal{A}_{2,m}\rangle+\vert\mathcal{A}_{3,m}\rangle\right)/2$
and $\vert\chi_{m,0}\rangle=\left(\vert\mathcal{A}_{1,m}\rangle-\vert\mathcal{A}_{3,m}\rangle\right)/\sqrt{2}$
associated with the $m$th unit cell. As shown in Eq.~(\ref{eq:3-edgestates}), then these edge states can be given as
\begin{align}
\vert\varPsi_{\pm}\rangle & =\stackrel[m=1]{M}{\sum}\lambda^{m}\left(\frac{\sigma_{A_{1,m}}^{+}\pm\sqrt{2}\sigma_{A_{2,m}}^{+}+\sigma_{A_{3,m}}^{+}}{2}\right)\vert G\rangle\, \nonumber \\
\vert\varPsi_{0}\rangle & =\stackrel[m=1]{M}{\sum}\lambda^{m}\left(\frac{\sigma_{A_{1,m}}^{+}-\sigma_{A_{3,m}}^{+}}{\sqrt{2}}\right)\vert G\rangle.
\end{align}
When $\left|\lambda\right|\ll1$, i.e., $v\ll w$, the edge states
are mainly localized at the left end of the chain, and when $\left|\lambda\right|\gg1$,
i.e., $v\gg w$, the edge states are mainly localized at the right end of the chain. In particular,
when $v=0$, these edge states are written as $\vert L_{\pm}\rangle=\vert\chi_{1,\pm}\rangle\vert gg\cdots g\rangle$ and $\vert L_{0}\rangle=\vert\chi_{1,0}\rangle\vert gg\cdots g\rangle$, supported by the transport qubits on the left end of the chain.
However, when $w=0$, the edge states are $\vert R_{\pm}\rangle=\vert gg\cdots g\rangle\vert\chi_{M,\pm}\rangle$ and
$\vert R_{0}\rangle=\vert gg\cdots g\rangle\vert\chi_{M,0}\rangle$,  supported by the transport qubits on the  right end of the chain.

\begin{figure}[htb]
\begin{centering}
\includegraphics[width=9cm]{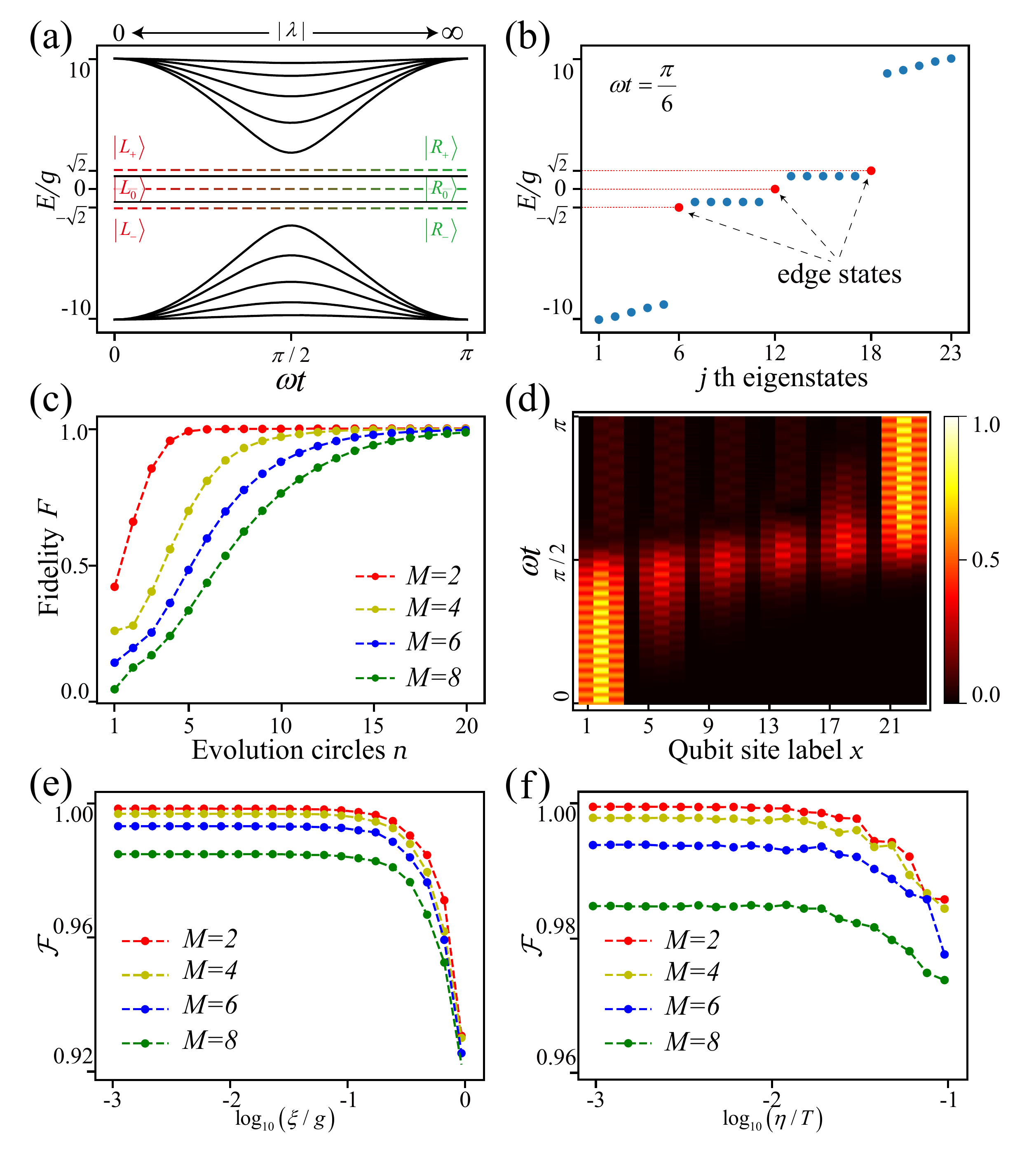}
\par\end{centering}
\caption{\label{Fig:3-qubit state transfer} $3$-qubit state transfer with extended SSH$4$ chain. The parameters are taken as $J=5g$ and $g/2\pi=10$MHz. (a) The energy spectrum of a qubit chain with 6 unit cells when $\omega t$
changes from $0$ to $\pi$. Edge states can be transferred from left edge (red)
to right edge (green) along the dashed energy levels. (b) Schematics for eigenstate distribution of the SSH$4$ chain in a topological nontrivial regime ($v\ll w$) when $\omega t=\pi/6$. Four bulk bands (blue dots), are divided by three edge states (red dots). (c) Target-state
occupation probabilities $F(nT)=\left|\langle\varPsi_{F}\vert\varPsi\left(nT\right)\rangle\right|$ as a function of the number of evolution circles $n$. Each point represents a complete adiabatic evolution from $t=0$ to $t=\pi /\omega$. Various colors from red to green represent different lengths of the chain, where $M=2,4,6,8$. (d) Time evolution of the whole qubit-chain
state with initial state prepared to a three-qubit W state $\left(\vert \mathcal{A}_{1,1}\rangle+\vert \mathcal{A}_{2,1}\rangle+\vert \mathcal{A}_{3,1}\rangle\right)/\sqrt{3}$ when the total evolution time is $20$ evolution circles $n$. The color from dark-red to bright-yellow represents the population distribution of the state on each qubit site. (e) The average fidelities of 3-qubit W-state transfer with different coupling disorder strength. $M=2,4,6,8$ for each curve, separately. (f) The average fidelities of $3$-qubit W-state
transfer with different execution time disorder.}
\end{figure}

As discussed above, we can change the coupling constants slowly as $v=J\left[1-\cos\left(\omega t\right)\right]$ and $w=J\left[1+\cos\left(\omega t\right)\right]$, then the edge states of the system will adiabatically evolve from the left end to the right end of the chain. As shown in Fig.~\ref{Fig:3-qubit state transfer}(a), using a qubit chain with $M=6$ unit cells (i.e., $23$ qubits) as an example,
we plot the variations of the instantaneous eigeneneries from $t=0$ to $\pi/\omega$. We find that there are four bands of bulk states, separated by three topological edge states. The bulk bands outside of the edge states are not degenerate, however, each bulk band inside the gap of the edge states has five-fold degenerate. This has been further illustrated in Fig.~\ref{Fig:3-qubit state transfer}(b) by arranging $23$ eigenenstates by the corresponding eigenenergies (from the lowest to the highest), when the time $t$ is taken as $t=\pi/6\omega$. Figure~\ref{Fig:3-qubit state transfer}(a) also shows if the system is prepared to one of the left edge states ($\vert L_{\pm}\rangle$ and $\vert L_{0}\rangle$) at $t=0$, then the state will evolve to the corresponding right edge state ($\vert R_{\pm}\rangle$ or $\vert R_{0}\rangle$) when $t_{f}=\pi/\omega$.

More generally, if the initial state is prepared to an arbitrary 3-qubit entangled
state at the left end of the chain as
\begin{equation}
 \vert\varPsi_{\rm in}\rangle=\left(\alpha\sigma_{A_{1,1}}^{+}+\beta\sigma_{A_{2,1}}^{+}+\gamma\sigma_{A_{3,1}}^{+}\right)\vert G\rangle,
\end{equation}
which can be rewritten as
\begin{equation}
\vert\varPsi_{\rm in}\rangle=\frac{\alpha+\sqrt{2}\beta+\gamma}{2}\vert L_{+}\rangle+\frac{\alpha-\sqrt{2}\beta+\gamma}{2}\vert L_{-}\rangle+\frac{\alpha-\gamma}{\sqrt{2}}\vert L_{0}\rangle,
\end{equation}
by using the left edge states, then the state will adiabatically evolve to
\begin{eqnarray}
\vert\varPsi\left(t\right)\rangle & = & \frac{\alpha+\sqrt{2}\beta+\gamma}{2}\vert\varPsi_{+}\left(t\right)\rangle e^{-i\int_{0}^{t}E_{+}dt'}\nonumber \\
 & + & \frac{\alpha-\sqrt{2}\beta+\gamma}{2}\vert\varPsi_{-}\left(t\right)\rangle e^{-i\int_{0}^{t}E_{-}dt'}\nonumber \\
 & + & \frac{\alpha-\gamma}{\sqrt{2}}\vert\varPsi_{0}\left(t\right)\rangle e^{-i\int_{0}^{t}E_{0}dt'},
\end{eqnarray}
at the moment $t$. As we learn from the case of the two-qubit state transfer, $E_{\pm}=\pm\sqrt{2}g$ and $E_{0}=0$ here are still constant during the adiabatic protocol, thus the final state at $t_{f}=\pi/\omega$ is
\begin{eqnarray}
\left\vert\varPsi_{f}\right\rangle & = & \frac{\alpha+\sqrt{2}\beta+\gamma}{2}\vert R_{+}\rangle e^{-i2\pi\frac{g}{\sqrt{2}w}}\nonumber \\
 & + & \frac{\alpha-\sqrt{2}\beta+\gamma}{2}\vert R_{-}\rangle e^{i2\pi\frac{g}{\sqrt{2}w}}\nonumber \\
 & + & \frac{\alpha-\gamma}{\sqrt{2}}\vert R_{0}\rangle.
\end{eqnarray}
Again, let us consider the evolution time to be exact integral multiples
of the dynamical period $T=2\pi/\sqrt{2}g$, i.e., $t_{f}/T=g/\sqrt{2}\omega=n$ ($n\gg1$), then the final state becomes
\begin{equation}
\vert\varPsi_{F}\rangle=\left(\alpha\sigma_{A_{1,M}}^{+}+\beta\sigma_{A_{2,M}}^{+}+\gamma\sigma_{A_{3,M}}^{+}\right)\vert G\rangle.
\end{equation}
Clearly then, as time $t$ changes from $0$ to $\pi/\omega$, arbitrary $3$-qubit entangled state can be transported from the left end to the right end of the chain.

Same as the case of $2$-qubit state transfer, we choose different numbers of evolution circles to examine the adiabaticity. In Fig.~\ref{Fig:3-qubit state transfer}(c),  target-state occupation probabilities $F(t)=\left|\langle\varPsi_{F}\vert\varPsi\left(t\right)\rangle\right|$ are plotted as a function of the number of evolution circles when a W-state is transferred from the left edge to the right one for different lengths of the chain. We find that the evolution circles to achieve high fidelity increase with the length $M$ of the chain when the qubit number in the unit cell is given.
For example, Fig.~\ref{Fig:3-qubit state transfer}(c) shows that $20$ evolving circles are required to achieve fidelity one when $M=8$, however $5$ evolving circles are enough to achieve fidelity one when $M=2$.  For a specific case shown in Fig.~\ref{Fig:3-qubit state transfer}(d), the W state is shown to be transferring from the left to the right with a rapid oscillation within the execution time (20 evolution circles). In Fig.~\ref{Fig:3-qubit state transfer}(e), we have also evaluated the average fidelities of the 3-qubit W-state transfer with different disorder
strengths $\xi$ for the coupling strengths. There is also a plateau at $\xi\in\left[10^{-3}g,10^{-1}g\right]$
for each qubit chain. However, different from the two-qubit transfer,
these plateaus are pinned at different values of fidelity with the increase of the qubit number.
We find that the average fidelity
for all the qubit chains considered here is far beyond $96\%$ so long as the disorder strength is $\xi<0.1g$. Meanwhile, the effect of the disorder of the time evolution is presented in Fig.~\ref{Fig:3-qubit state transfer}(f). The average fidelity $\mathcal{F}$ is also good enough for state transfer as the disorder strength $\eta$ is less than $0.01T$. Figs.~\ref{Fig:3-qubit state transfer}(e) and (f) clearly show that our proposal is also promising for transferring 3-qubit state along a long qubit chain.

\subsection{$\mathcal{N}$-qubit state transfer}

\begin{figure}[h]
\begin{centering}
\includegraphics[width=9cm]{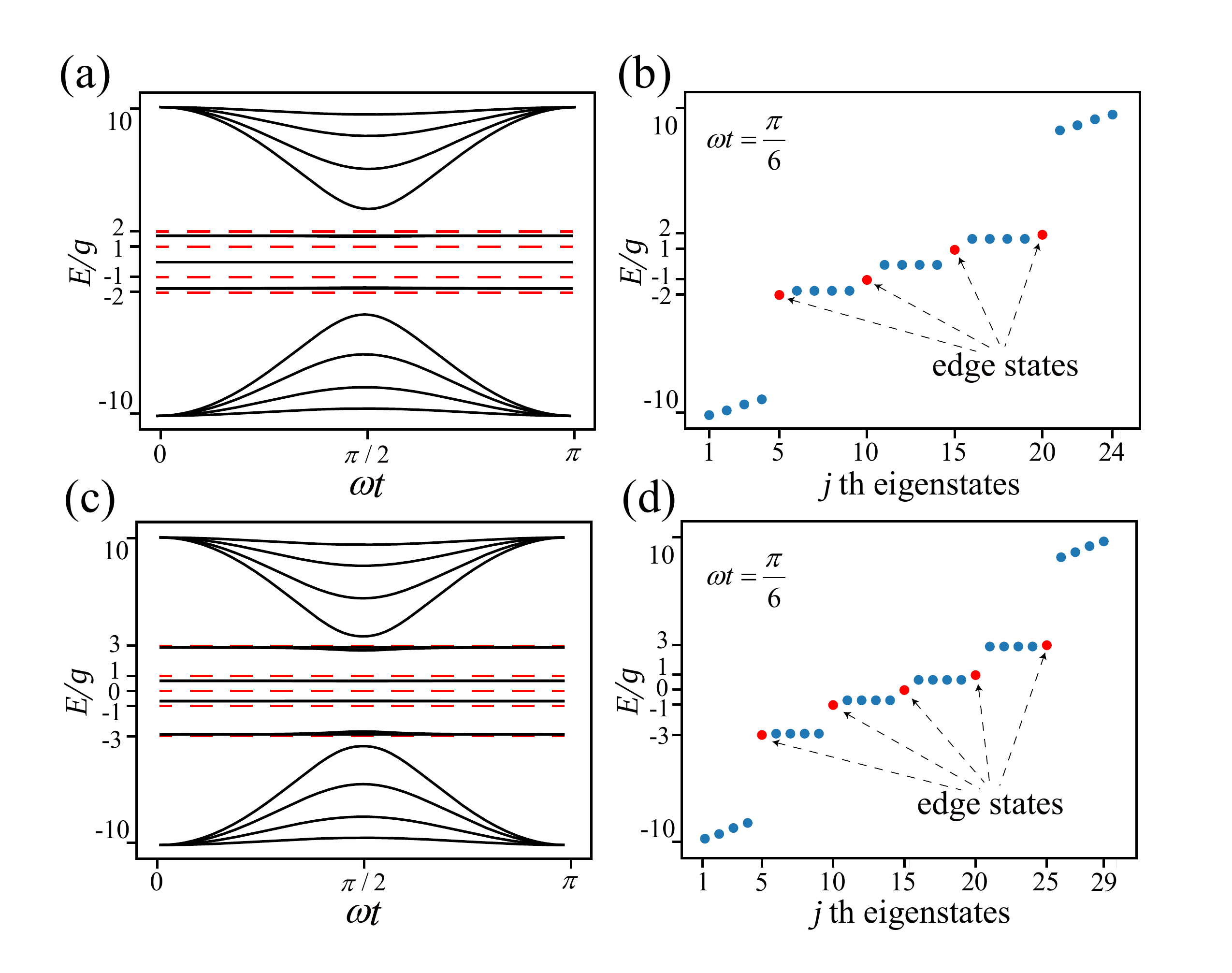}
\par\end{centering}
\caption{\label{Fig:4 and 5 qubit} (a) Spectrum of the qubit chain for 4-qubit state transfer. Here, $M=5$ and $\mathcal{L}=24$. The red-dashed lines represent the edge states and the black-solid lines represent the bulk states. (b) Schematics for eigenstates distribution corresponding to (a) in a topologically nontrivial regime ($v\ll w$) when $\omega t=\pi/6$. Four red dots denote four edge states. The blue dots denote the bulk bands. The four blue dots in line between two red dots denote a bulk band with $4$-fold degenerate. (c) Spectrum of the qubit chain for 5-qubit state transfer. Here, $M=5$ and $\mathcal{L}=29$. (d) Schematics for eigenstates distribution corresponding to (c) in a topologically nontrivial regime ($v\ll w$) when $\omega t=\pi/6$. Five red dots denote five edge states. The blue dots denote the bulk bands. The four dots in line between two red dots denote a bulk band with $4$-fold degenerate.}
\end{figure}

We have extended our protocol from 2-qubit state transfer to 3-qubit state transfer. Analogously, we can extend our proposal to $\mathcal{N}$-qubit state case. Similar to Eq.~(\ref{eq:psi}) for the two-qubit entangled state transfer, the edge states  corresponding to the Hamiltonian in Eq.~(\ref{eq:H-N}) do not occupy the mediated qubits $B$ and can be expanded as
\begin{equation}
	\vert\varPsi_{\rm edge}\rangle=\stackrel[m=1]{M}{\sum}\lambda^{m}\left(\chi_{1}\sigma_{A_{1,m}}^{+}+\cdots+\chi_{\mathcal{N}}\sigma_{A_{\mathcal{N},m}}^{+}\right)\vert G\rangle.\label{eq:20}
\end{equation}
Substituting Eq.~(\ref{eq:H-N}) and Eq.~(\ref{eq:20}) into the Schr\"odinger equation $H\vert\varPsi_{\rm edge}\rangle=E\vert\varPsi_{\rm edge}\rangle$, then we can have
\begin{align}
	E & \left(\chi_{1}\sigma_{A_{1,m}}^{+}+\chi_{2}\sigma_{A_{2,m}}^{+}+\cdots+\chi_{\mathcal{N}}\sigma_{A_{\mathcal{N},m}}^{+}\right)\vert G\rangle\nonumber \\
	& =  \left(g_{1}\chi_{2}\sigma_{A_{1,m}}^{+}+\cdots+g_{\mathcal{N}-1}\chi_{\mathcal{N}}\sigma_{A_{\mathcal{N}-1,m}}^{+}\right)\vert G\rangle\nonumber \\
	& +
	\left(g_{1}\chi_{1}\sigma_{A_{2,m}}^{+}+\cdots+g_{\mathcal{N}-1}\chi_{\mathcal{N}-1}\sigma_{A_{\mathcal{N},m}}^{+}\right)\vert G\rangle\nonumber \\
	& +  \chi_{1}\left(v\sigma_{B_m}^{+}+w\lambda\sigma_{B_m}^{+}\right)\vert G\rangle.\label{eq:21}
\end{align}
From Eq.~(\ref{eq:21}), it is straightforward to get $\lambda=-v/w$, and the coefficients
$\chi_{1}$, $\chi_{2},\cdots,\chi_{\mathcal{N}}$ satisfy the following equation
\begin{equation}
	\left(\begin{array}{ccccc}
		0 & g_{1} & 0 & 0 & 0\\
		g_{1} & 0 & g_{2} & 0 & 0\\
		0 & g_{2} & 0 & \ddots & 0\\
		0 & 0 & \ddots & \ddots & g_{\mathcal{N}-1}\\
		0 & 0 & 0 & g_{\mathcal{N}-1} & 0
	\end{array}\right)\left(\begin{array}{c}
		\chi_{1}\\
		\chi_{2}\\
		\chi_{3}\\
		\vdots\\
		\chi_{\mathcal{N}}
	\end{array}\right)=E\left(\begin{array}{c}
		\chi_{1}\\
		\chi_{2}\\
		\chi_{3}\\
		\vdots\\
		\chi_{\mathcal{N}}
	\end{array}\right).\label{eq:N-eigen}
\end{equation}
Solving this eigen-equation, we can obtain the eigenenergies of the edge
states as $E_{1},\cdots, E_{\mathcal{N}}$. The explicit eigenstates can also be derived, and then edge states can be obtained. Similar to Eq.~(\ref{eq:vw}), when the time $t$ is adiabatically changed from $0$ to $\pi/\omega$, the coupling constant $v$ ($w$) is changed from $2J$ ($0$) to $0$ ($2J$), and the edge states initially at the left end of the chain can be adiabatically transferred to the right end.  We know that an arbitrary $\mathcal{N}
$-qubit entangled state can be represented by these edge states, thus $\mathcal{N}
$-qubit entangled state can also be perfectly transferred from the left end to the right end of the chain if the total evolution time $t_{f}=\pi/\omega$ is exact multiple of all  the dynamical periods corresponding to all eigenenergies $E_{1},\cdots, E_{\mathcal{N}}$. That is, there must be a least common period of all dynamical periods $T_{1}=2\pi/E_{1}, \cdots, T_{\mathcal{N}}=2\pi/E_{\mathcal{N}} $ corresponding to the $\mathcal{N}$ edge states. This common period can only exist for specific set of parameters $\left\{ g_{1},\cdots,g_{\mathcal{N}-1}\right\}$.

It seems that the generalization is straightforward, however the main problem is how to
engineer the proper series of parameters $\left\{ g_{1},\cdots,g_{\mathcal{N}-1}\right\}$ to get a common period.
It is highly challenging to find a general solution for any $\mathcal{N}$, because the solution varies case by case.
However, for given qubit states, we can always engineer the coupling constants between the transport qubits such that these states can be transferred through the extended SSH chain.  Below,
two examples for $\mathcal{N}=4$ and $\mathcal{N}=5$ are further shown in Fig.~\ref{Fig:4 and 5 qubit}.
At $\mathcal{N}=4$, the parameters $\left\{ g_{1},g_{2},g_{3}\right\} $
can be set as $\left\{ \sqrt{2}g,g,\sqrt{2}g\right\} $ and the corresponding
eigenenergies are $\left\{ 2g,g,-g,-2g\right\} $. In Figs.~\ref{Fig:4 and 5 qubit}(a) and (b), we show the spectrum of the qubit chain for $4$-qubit state transfer with a number of $24$ qubits. Five bulk bands are divided by four edge states, and the three bulk bands between edge states are $4$-fold degenerate. The least common oscillation period of these edge states is $2\pi/g$, which can be taken as the ideal time evolution circle. At $\mathcal{N}=5$, the parameters
$\left\{ g_{1},g_{2},g_{3},g_{4}\right\} $ can be set as $\left\{ g,2g,2g,g\right\} $
and the corresponding eigenenergies are $\left\{ 3g,g,0,-g,-3g\right\} $. The spectrum of the qubit chain in this case is plotted in Figs.~\ref{Fig:4 and 5 qubit}(c) and (d) for the total number $29$ of the qubits in the chain. Six bulk bands are divided by five edge states, and the four bulk bands between edge states are $4$-fold degenerate.
The least common period is also $2\pi/g$. Similarly, for the case of $\mathcal{N}$-qubit state transfer, we can find a proper set of parameters to get a common period as the time evolution circle. By use of such time evolution cycles, the involved edge states will not suffer from mutual dynamical phase differences at the end of our adiabatic protocol,
and hence an arbitrary $\mathcal{N}$-qubit
entangled state can be transferred from the left end to the right end.

\section{discussions}

\subsection{General discussions}

We have proposed a QST approach along an extended SSH qubit chain. However, several issues
should be further discussed. First, we consider only the neighbor-couplings and on-site potentials are not included
as shown in Eq.~\eqref{eq:H}. This is because the on-site potentials only result in a global dynamical phase if all the qubits are tuned to resonate with each other, and thus this global phase can be dropped out. Second, we note that the gap between the edge states and the bulk states decreases with the increase in length of the qubit chain. This indicates that the adiabatic condition is highly demanded for very large systems. However, as shown in Fig.~\ref{Fig:3-qubit state transfer}(c), for an extended SSH4 chain consisting of $8$ unit cells, the adiabatic condition is still well met for an execution time of only 20 evolution circles. When the length of the chain becomes much larger, one possible solution is hinted in Ref.~\citep{tan2020high}. That is,
our QST protocol realized by one-step adiabatic evolution can be decomposed into multi-step process to achieve a better performance.

To verify the topological protection of our QST protocol, one may compare our approach with QST protocols independent of topology.  Using mirror symmetry, one can think of an obvious non-topological QST protocol in a qubit chain as follows. At the beginning, the leftmost qubit is decoupled from other part of the chain, and at the end the couplings of the qubit chain is slowly changed to the reversed form, i.e., the rightmost qubit is decoupled with other qubits. With this adiabatic process, the quantum state initially prepared to the leftmost qubit can be transferred from the left end to the right end. Ref.~\citep{palaiodimopoulos2021fast} has shown that such topologically-unrelated QST protocol is not robust to disorder. Hence, our topology-based protocol does have an advantage, accounting for the fairly good fidelity presented above in the presence of disorder.

\subsection{Discussions for implementations using superconducting qubit circuits}

In principle, our proposal can be implemented in various platforms, e.g., cold atoms~\citep{du2016experimental}, trapped ions~\citep{harty2014high}, or coupled waveguides~\citep{shen2020acoustic,chen2021landau}. However, with the significant development in recent years, superconducting qubit circuits, e.g., transmon or Xmon qubits~\citep{chen2014qubit}, seem to be a more promising platform~\citep{PhysRevA.76.042319}. Also, a fast and high-fidelity transfer of arbitrary single-qubit state in a chain of superconducting qubits has been achieved in experiment recently~\citep{PhysRevApplied.10.054009}.

The good scalability and flexible tunability for the couplings make our extended SSH chain easy to be realized by the superconducting qubit circuits. In particular, the coupling strengths can be tuned from topologically trivial to non-trivial regime. Thus topological and non-topological phenomena can be studied in one quantum circuits. Moreover,  the coherent time is a very important for realizing our proposal. As shown in Appendix~\ref{sec:Xmon-qubut-chain}, the coupling coefficient $g$ can be chosen as $g/2\pi=10$Mhz by using the parameters of current superconducting qubit circuits for realizing our proposal. Therefore, for two-qubit state transfer, one evolution circle can be $T=2\pi/g=0.1\mu s$. The total evolution time is 10 evolution circles, i.e., $t_{f}=1\mu s$. For three-qubit state transfer, one evolution circle can be $T=2\pi/\sqrt{2}g=0.0707\mu s$. The total evolution time is 20 evolution circles, i.e., $t_{f}=1.414\mu s$. In superconducting qubit circuits,  the coherence time of single qubit is about $10\sim100\mu s$ ~\citep{PhysRevLett.107.240501,novikov2016raman}, which is much longer than the adiabatic time in our proposal. Meanwhile, steady topological edge states have already been observed in superconducting qubit circuits, which can last more than $1\mu s$~\citep{cai2019observation}. Therefore, the decoherence effect is not expected to be troublesome in our protocol.

To make sure that the adiabatic evolution time for our proposal is indeed integral multiple of the evolution circle, one needs to measure the exact value of $g$, which is equivalent to determining the eigenenergy of edge states. This can be achieved by the reflection spectrum of a weak probe light through a waveguide coupled to the extended SSH chain (see Appendix~\ref{sec:Energy-spectrum-with}). The reflection peaks of the input weak signal can reveal the energy spectrum of the qubit chain with appropriate parameters, and the value of coupling $g$ can be obtained with the energy shift between the edge states.

\section{Conclusion \label{sec:discussions-and-conclution}}

In summary, we have proposed an experimentally feasible approach for transferring
arbitrary entangled state through an extended SSH chain.
The entangled states are encoded in the edge states of a class of extended SSH chains,
and then they are transported via an adiabatic protocol. Due to the topological
protection of the edge states, our protocol is robust against the the temporal
noise caused by the imperfection in the control field. We have numerically confirmed this robustness
against two kinds of disorders, i.e, the coupling strength disorder and the execution time disorder. Compared with most contemporary studies realizing QST in qubit chains, this work represents an exciting advance that a general scenario is proposed to achieve QST of arbitrary $\mathcal{N}$-qubit entangled state. Our proposal is easy to be realized by using superconducting qubit circuits and the parameters required in our protocol are estimated according to the recent experiments. Given the feasibility and tunability of the proposed protocal, our idea can also be extended for realizing QST in two dimensional quantum networks.

\section{ACKNOWLEDGMENTS}
We thank Wei Nie for helpful discussions. J.G. acknowledges funding support by the Singapore NRF Grant No. NRF-NRFI2017-04 (WBS No. R-144-000-378- 281).  Y.X.L. is supported by the Key-Area Research and Development Program of GuangDong Province under Grant No. 2018B030326001, the National Basic Research Program (973) of China under Grant No. 2017YFA0304304, and the NSFC under Grant No. 11874037.

\appendix

\section{Xmon qubit chain with tunable couplings\label{sec:Xmon-qubut-chain}}

For our state transfer protocol, the feasibility comes from the precise tuning of the coupling between qubits in the topological qubit chain. We manage to realize such setup with superconducting qubits.
A superconducting qubit chain with tunable couplings is presented in Fig.~\ref{Fig:Model scheme}(a).
The details of the coupler circuit are schematically shown in Fig.~\ref{Fig:Model scheme}(b), where two Xmon qubits are coupled by a tunable Josephson junction coupler.  Such coupling scheme has been experimentally realized
in Ref.~\citep{chen2014qubit} and theoretically analyzed in Ref.~\citep{geller2015tunable}.

\begin{figure}[htp]
	\begin{centering}
		\includegraphics[width=9cm]{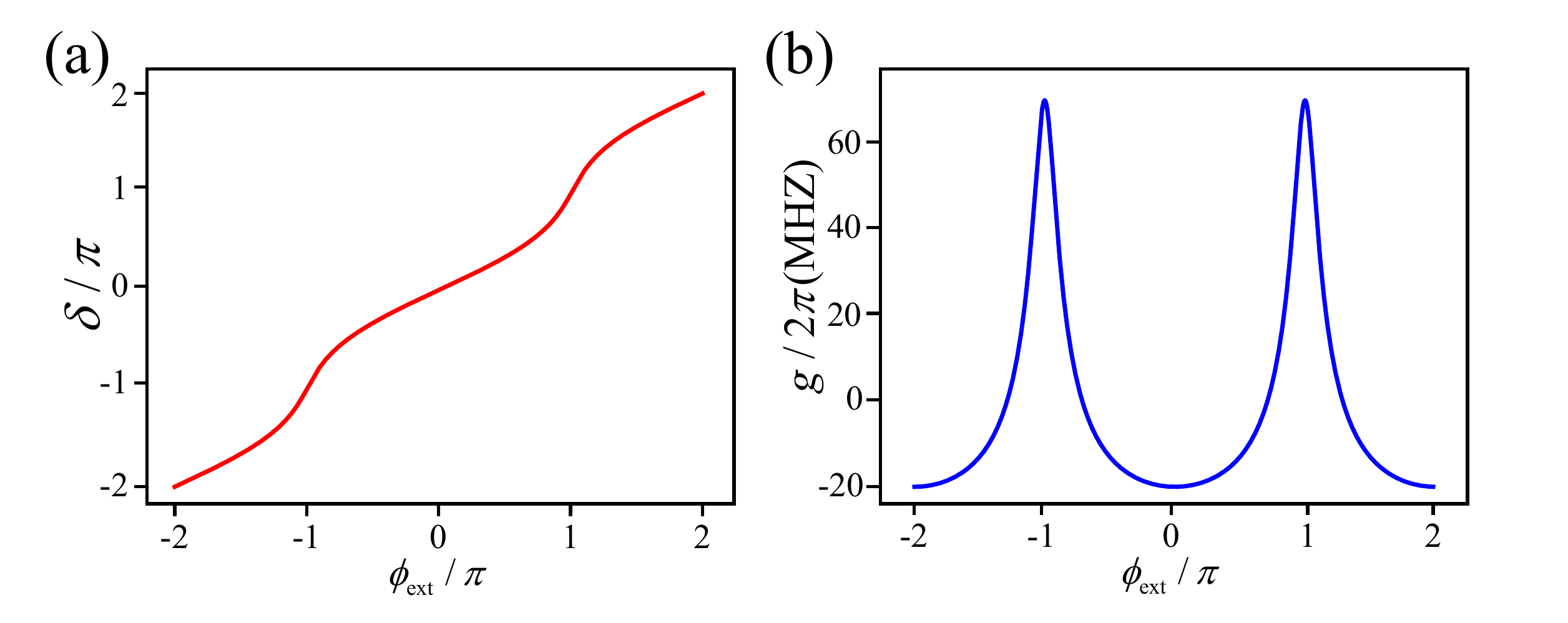}
		\par\end{centering}
	\caption{\label{Fig:Xmon} (a) The change of $\delta$ versus $\phi_{\text{ext}}$. The parameters we use here are $L_{\rm g}=300$ pH and $L_{\rm T}=1$ nH. (b) Coupling strength varies with  $\phi_{\text{ext}}$. The parameters are  $L_{\rm J}=8$ nH and $\omega_{q}=5.5$ GHz. Other parameters are the same as above.}
\end{figure}

The junction connecting two qubits provides a tunable effective inductance $L_{\text{eff}}$ to tune the coupling, and a magnetic flux bias $\Phi_{\rm ext}$ is used to tune the coupler's effective inductance $L_{\text{eff}}=L_{\text{T}}/\cos \delta$~\citep{geller2015tunable}, where $\delta$ is the phase difference across the coupler. For the coupler loop, we have $\Phi=(\Phi_{0}/2\pi)\delta$, where $\Phi$ is the total magnetic flux and $\Phi_{0}=h/2e$. $L_{\text{T}}=\Phi_{0}/2\pi I_{c}$ is the zero-bias inductances of the Josephson coupler and $I_{c}$ is the critical current of the coupler junction. The circuit flux has the relation $\Phi=\Phi_{\rm ext}-2L_{\text{g}}I_{c}\sin \delta$ and therefore we can have
\begin{equation}
	\phi_{\rm ext} = \delta+\left(\frac{2L_{\text{\text{g}}}}{L_{\text{T}}}\right)\sin\delta \label{eq:delta-phi}
\end{equation}
where $\phi_{\rm ext}=2\pi\Phi_{\rm ext}/\Phi_{0}$.
In the
the weakly coupled limit the effective coupling strength is approximately~\citep{chen2014qubit}
\begin{equation}
	g=-\frac{\omega_{q}}{2}\frac{M}{L_{\textrm{J}}+L_{\textrm{g}}},
\end{equation}
where $M=L_{\textrm{g}}^{2}/\left(2L_{\textrm{g}}+L_{\textrm{eff}}\right)$ is the mutual inductance and $\omega_{q}$ is the qubit frequency. Thus the coupling strength can be finally given by~\citep{geller2015tunable}
\begin{eqnarray}
g & = & -\frac{L_{\text{g}}^{2}\cos\delta}{2\left(L_{\text{J}}+L_{\text{g}}\right)\left(L_{\text{T}}+2L_{\text{g}}\cos\delta\right)}\omega_{q}.
\end{eqnarray}

In Fig.~\ref{Fig:Xmon}(a), we present the relation between $\Phi_{\rm ext}$ and $\delta$. The phase $\delta$ can be tuned from $-2\pi$ to $2\pi$ when the external magnetic flux $\Phi_{\text{ext}}$ is continuously changed. In Fig.~\ref{Fig:Xmon}(b), we show how the coupling between qubits can be tuned with $\Phi_{\text{ext}}$. As $\phi_{\text{ext}}$ is changing from $-2\pi$ to $2\pi$, the coupling can be tuned from $-20$MHZ to $60$MHZ, which meet the requirement of the system parameters to realize our proposal.

\section{A straightforward diagram for exact solution of the edge states\label{sec:A-straightforward-diagram}}

As for standard SSH model, there are only approximated solutions for hybridized
edge states~\citep{asboth2016short}, but for our imperfect SSH model,
one can obtain the exact solution of the edge states~\citep{mei2018robust}.
To illustrate this process straightforwardly, let us rewrite the
qubit array with specific set of basis. As shown in Fig.~\ref{Fig:Model scheme},
we here analyze $2$-qubit state transfer, and the
formula can be extended to $3$-qubit and $\mathcal{N}$-qubit
cases.

\begin{figure}[htp]
\begin{centering}
\includegraphics[width=9cm]{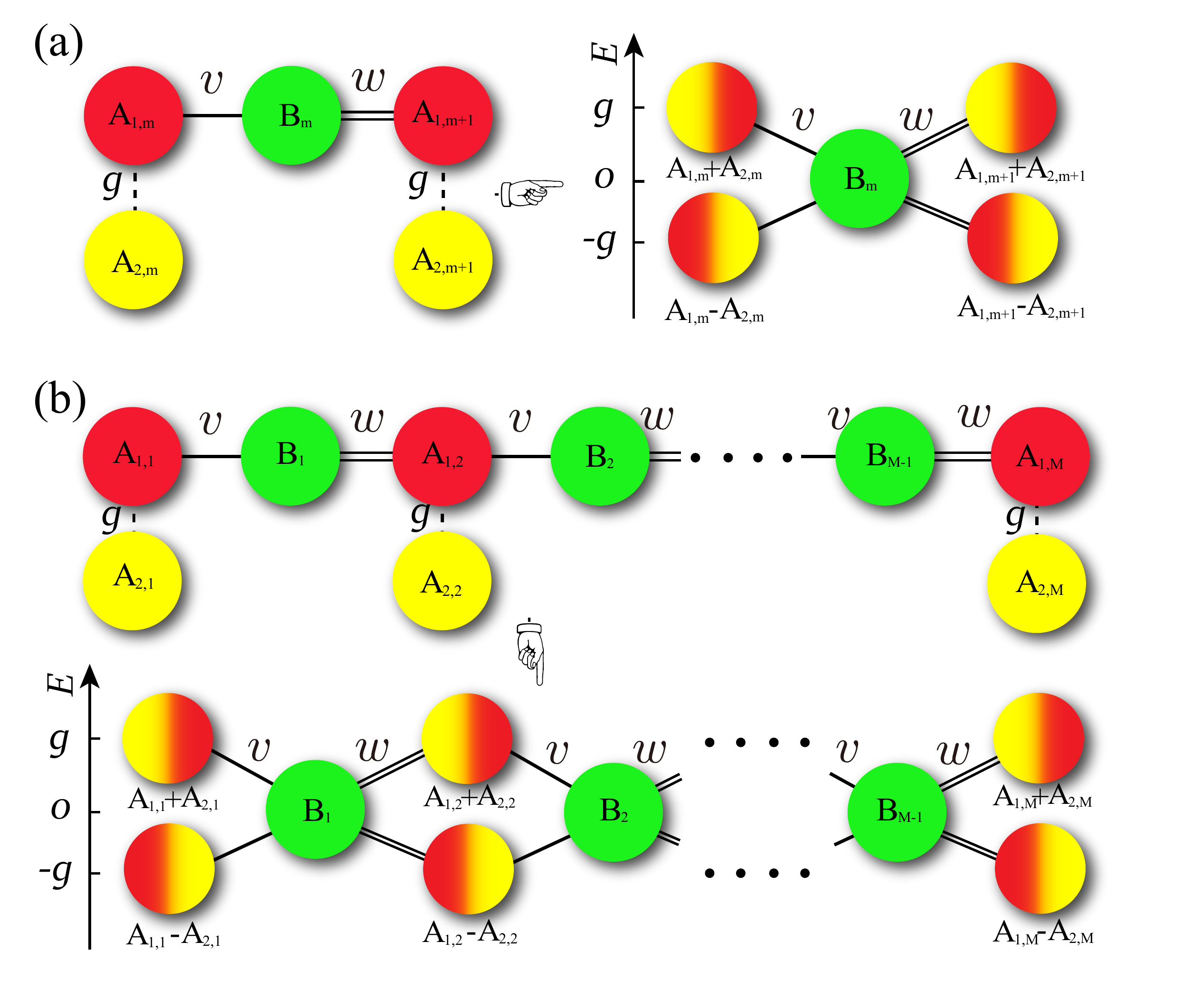}
\par\end{centering}
\caption{\label{Fig:diagram-renormal}Schematic diagram of the renormalization process
on qubit array with orthogonal basis of transport qubits. (a) For
each unit cell of the qubit chain, the renormalization leads to two on-site
potentials as $g$ and $-g$, due to the coupling strength $g$ between
the transport qubits $A_{1,m}$ and $A_{2,m}$. (b) Renormalization of the whole qubit
array. All the transport qubits are decoupled with each other, and
the hopping amplitudes are staggered through the mediated qubits $B_{m}$.
The qubit array now can be distinguished into two branches by the
on-site energy: the upper branch with the hybrid qubit $A_{1,m}+A_{2,m}$ and the lower branch with the hybrid qubit $A_{1,m}-A_{2,m}$.}
\end{figure}

For 2-qubit state transfer, the Hamiltonian of the qubit chain is
\begin{align}
H & =  \stackrel[m=1]{M-1}{\sum}\left(v\sigma_{A_{1,m}}^{+}\sigma_{B_m}^{-}+w\sigma_{A_{1,m+1}}^{+}\sigma_{B_m}^{-}+\textrm{H.c.}\right)\nonumber \\
 & +  \stackrel[m=1]{M}{\sum}\left(g\sigma_{A_{1,m}}^{+}\sigma_{A_{2,m}}^{-}+\textrm{H.c.}\right),\label{eq:diagram-H}
\end{align}
same as Eq.~\eqref{eq:H}. As shown in Fig. \ref{Fig:diagram-renormal}(a), for
a single unit cell of the qubit chain, the Hamiltonian is
\begin{align}
H_{\text{cell}} & =  v\sigma_{A_{1,m}}^{+}\sigma_{B_m}^{-}+w\sigma_{A_{1,m+1}}^{+}\sigma_{B_m}^{-}+\textrm{H.c.}\nonumber \\
 & +  g\sigma_{A_{1,m}}^{+}\sigma_{A_{2,m}}^{-}+g\sigma_{A_{1,m+1}}^{+}\sigma_{A_{2,m+1}}^{-}+\textrm{H.c.}.\label{eq:diagram-cell}
\end{align}
In the single-excitation subspace, the Hamiltonian in Eq.\eqref{eq:diagram-cell}
can be rewritten as
\begin{align}
H_{\text{cell}} & = v\vert\mathcal{A}_{1,m}\rangle\langle\mathcal{B}_{m}\vert+w\vert\mathcal{A}_{1,m+1}\rangle\langle\mathcal{B}_{m}\vert+\textrm{H.c.}\nonumber \\
 & + g\vert\mathcal{A}_{1,m}\rangle\langle\mathcal{A}_{2,m}\vert+g\vert\mathcal{A}_{1,m+1}\rangle\langle\mathcal{A}_{2,m+1}\vert+\textrm{H.c.}.\label{eq:diagram-cell-1}
\end{align}
The term of $g\vert\mathcal{A}_{1,m}\rangle\langle\mathcal{A}_{2,m}\vert+\textrm{H.c.}$
in matrix form is $\left(\begin{array}{cc}
0 & g\\
g & 0
\end{array}\right)$, and can be renomorlized as $\left(\begin{array}{cc}
g & 0\\
0 & -g
\end{array}\right)$ with basis of $\vert\chi_{m,+}\rangle=\left(\vert\mathcal{A}_{1,m}\rangle+\vert\mathcal{A}_{2,m}\rangle\right)/\sqrt{2}$
 and $\vert\chi_{m,-}\rangle=\left(\vert\mathcal{A}_{1,m}\rangle-\vert\mathcal{A}_{2,m}\rangle\right)/\sqrt{2}$.
In the basis \{$\vert\mathcal{\chi}_{m,+}\rangle,\vert\mathcal{\chi}_{m,-}\rangle,\vert\mathcal{B}_{m}\rangle$\}, the Hamiltonian $H_{\text{cell}}$ of the single unit cell can be rewritten as
\begin{eqnarray}
H_{\text{cell}}^{R} & = & v\frac{\vert\chi_{m,+}\rangle+\vert\chi_{m,-}\rangle}{\sqrt{2}}\langle\mathcal{B}_{m}\vert+\textrm{H.c.}\nonumber \\
 & + & w\frac{\vert\chi_{m+1,+}\rangle+\vert\chi_{m+1,-}\rangle}{\sqrt{2}}\langle\mathcal{B}_{m}\vert+\textrm{H.c.}\nonumber \\
 & + & g\vert\chi_{m+1,+}\rangle\langle\chi_{m+1,+}\vert-g\vert\chi_{m+1,-}\rangle\langle\chi_{m+1,-}\vert\nonumber \\
 & + & g\vert\chi_{m,+}\rangle\langle\chi_{m,+}\vert-g\vert\chi_{m,-}\rangle\langle\chi_{m,-}\vert.\label{eq:branched-cell-2}
\end{eqnarray}
The coupling between $A_{1,1}$ and $A_{2,1}$ in the unit cell leads to
two on-site potentials $g$ and $-g$. The total Hamiltonian of the
qubit chain in such new basis is
\begin{eqnarray}
H^{R} & = & \stackrel[m=1]{M-1}{\sum}\left(v\frac{\vert\chi_{m,+}\rangle+\vert\chi_{m,-}\rangle}{\sqrt{2}}\langle\mathcal{B}_{m}\vert+\textrm{H.c.}\right)\nonumber \\
 & + & \stackrel[m=1]{M-1}{\sum}\left(w\frac{\vert\chi_{m+1,+}\rangle+\vert\chi_{m+1,-}\rangle}{\sqrt{2}}\langle\mathcal{B}_{m}\vert+\textrm{H.c.}\right)\nonumber \\
 & + & \stackrel[m=1]{M}{\sum}g\left(\vert\chi_{m,+}\rangle\langle\chi_{m,+}\vert-\vert\chi_{m,-}\rangle\langle\chi_{m,-}\vert\right).\label{eq:diagram-total}
\end{eqnarray}
That is, as shown in Fig.~\ref{Fig:diagram-renormal}(b), the renormalized qubit chain
can be divided into two different branches by on-site potential.
The upper branch's transport qubits now are $A_{1,m}+A_{2,m}$ and lower branch's
transport qubits are $A_{1,m}-A_{2,m}$. Two branches have the same mediated qubits
$B_{m}$. The edge states for the upper and lower branches can be written
as
\begin{align}
\vert\varPsi_{\pm}\rangle & =\stackrel[m=1]{M}{\sum}\lambda^{m}\vert\chi_{m,\pm}\rangle\nonumber \\
 & =\stackrel[m=1]{M}{\sum}\lambda^{m}\left(\frac{\vert\mathcal{A}_{1,m}\rangle\pm\vert\mathcal{A}_{2,m}\rangle}{\sqrt{2}}\right),\label{eq:diagram-edge}
\end{align}
which is the same as Eq.~\eqref{eq:psi-1}. The extension of this analysis for $3$-qubit
and $\mathcal{N}$-qubit state transfer is straightforward, thus
we will not elaborate here.

\section{Topological invariant for our extended SSH model\label{sec:Topological-invariant}}

\begin{figure}[htb]
\begin{centering}
\includegraphics[width=9cm]{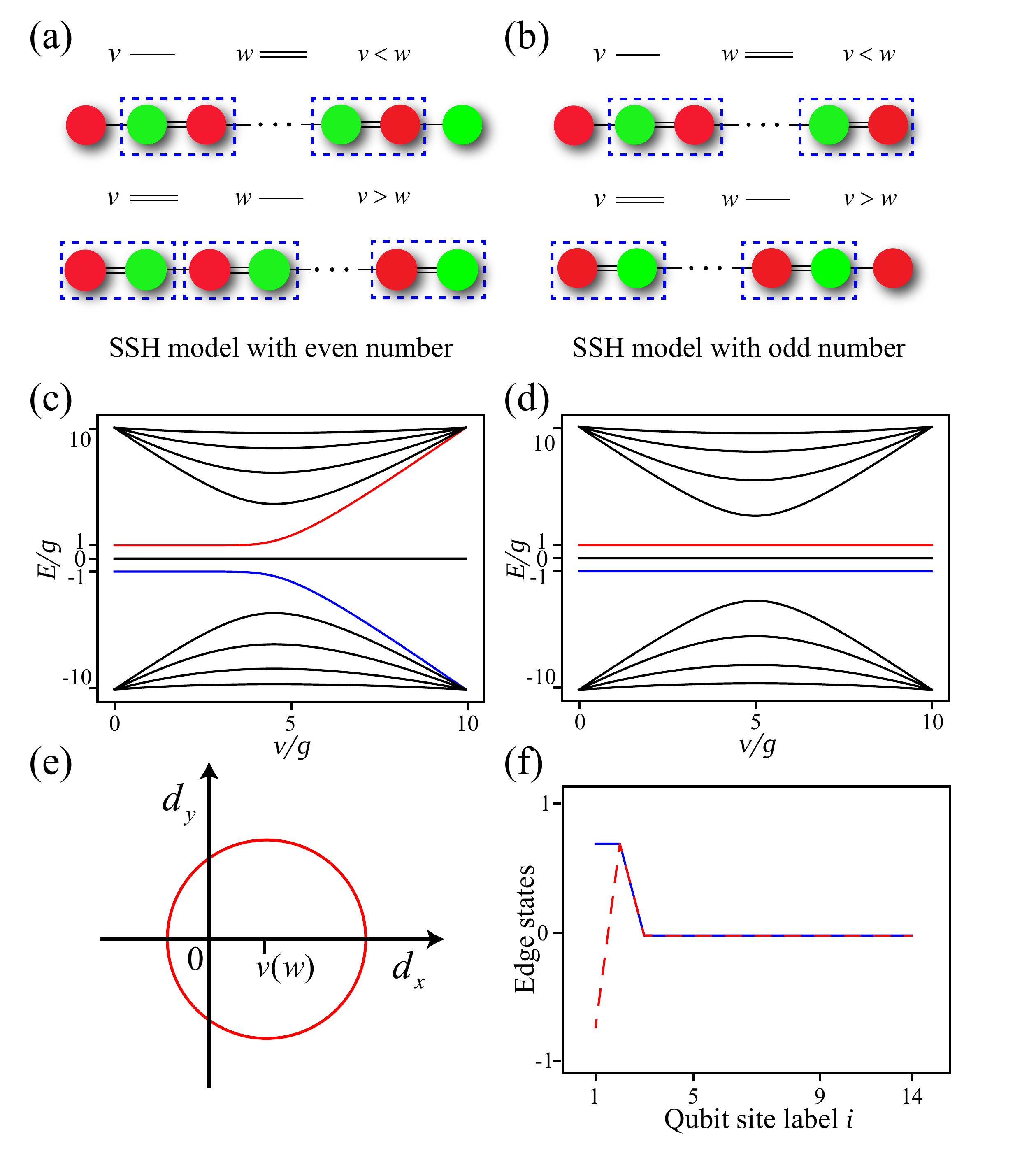}
\par\end{centering}
\caption{\label{Fig:diagram-renormal-1} Topological invariant and edge states for our
extended SSH model. (a) SSH model with an even number of qubits. When
$v<w$, the qubits at two end are insulated and edge states appear.
When $v>w$, all qubits are included in bulk states and edge states
disappear. (b) SSH model with an odd number of qubits. we can see
the system always has an edge state whether if $v<w$ or $v>w$. (c)
Energy spectrum for extended SSH model with complete $3M$ qubits. Two edge
states are denoted with blue- and red- lines. Topological phase transition
occurs with the increasing of $v$, and edge states join the bulk
band when $v\gg w$. (d) Energy spectrum for our extended SSH model with
$3M-1$ qubits, i.e., lacking of one qubit at the right end of the
chain. The system always has two edge states with the variation of
couplings. (e) Winding number in $d_{x}$-$d_{y}$ plane. When $v<w$ ($v>w$),
the centre of the integral circle is located at $v$ ($w$) on $d_{x}$ axis. (f) Edge states for our extended SSH model, i.e.,
$\left(\vert\mathcal{A}_{1,1}\rangle\pm\vert\mathcal{A}_{2,1}\rangle\right)/\sqrt{2}$.}
\end{figure}

In the standard SSH model with an even number of qubits, edge states
appear when the coupling strength at the edge is weaker than the coupling
at bulk. As shown in Fig.~\ref{Fig:diagram-renormal-1}(a), if $v<w$, the qubits
at two ends are isolated from bulk and form two edges. If $v>w$,
all qubits are included in bulks. For our extended SSH$3$ model, with
$g=0$, sublattices $A_{1,m}$ and $B_{m}$ are decoupled from $A_{2,m}$, forming the SSH model
with an odd number of qubits. The Hamiltonian of such SSH model is ($A_{1,m}$ is redefined as $A_{m}$)
\begin{align}
H_{\text{SSH}} & = \stackrel[m=1]{M-1}{\sum}\left(v\sigma_{A_{m}}^{+}\sigma_{B_m}^{-}+w\sigma_{A_{m+1}}^{+}\sigma_{B_m}^{-}+\text{H.c.}\right).
\end{align}
As shown in Fig.~\ref{Fig:diagram-renormal-1}(b), if $v<w$, the weaker coupling strength
is $v$ at the left edge. Otherwise, if $w<v$, the weaker coupling
strength is $w$ at the right edge. 

For a standard SSH model which can be obtained by setting $g=0$ of the SSH$3$ model, edge states
are supported by the topological invariant defined in the phase space. We first assume $v<w$ and the unit cell is $(A_{m},B_{m})$. Due to the translation invariance of the bulk, we can make Fourier
transforms to the vectors $\vert\mathcal{A}_{m}\rangle$ and $\vert\mathcal{B}_{m}\rangle$
as
\begin{equation}
\vert\mathcal{A}_{k}\rangle=\frac{1}{\sqrt{M}}\stackrel[m=1]{M}{\sum}e^{imk}\vert\mathcal{A}_{m}\rangle,
\end{equation}
\begin{equation}
\vert\mathcal{B}_{k}\rangle=\frac{1}{\sqrt{M}}\stackrel[m=1]{M}{\sum}e^{imk}\vert\mathcal{B}_{m}\rangle,
\end{equation}
for $k\in\left\{ \delta_{k},2\delta_{k},\cdots,M\delta_{k}\right\} $
with $\delta_{k}=2\pi/M$. Here $k$ is the wavenumber of
the first Brillouin zone~\citep{asboth2016short}. The bulk momentum-space
Hamiltonian $H\left(k\right)$ is defined as
\begin{equation}
H\left(k\right)=\underset{\imath,\jmath\in\left\{ \mathcal{A}_{k},\mathcal{B}_{k}\right\} }{\sum}\langle\imath\vert H_{\text{SSH}}\vert\jmath\rangle\vert\imath\rangle\langle\jmath\vert.\label{eq:Hk}
\end{equation}
By choosing a fixed momentum $k$,
we can get the matrix form of the bulk momentum-space Hamiltonian
$h\left(k\right)$ as
\begin{equation}
h\left(k\right)=\left(\begin{array}{cc}
0 & v+we^{-ik}\\
v+we^{ik} & 0
\end{array}\right).\label{eq:Hk-1}
\end{equation}
Thus the total Hamiltonian can be written as $H\left(k\right)=\sum_{k}\varPsi_{k}^{\dagger}h\left(k\right)\varPsi_{k}$,
with $\varPsi_{k}^{\dagger}=\left(\vert\mathcal{A}_{k}\rangle,\vert\mathcal{B}_{k}\rangle\right)$
and we have
\begin{equation}
h\left(k\right)=d_{x}\left(k\right)\sigma_{x}+d_{y}\left(k\right)\sigma_{y},
\end{equation}
where $d_{x}\left(k\right)=v+w\cos k$, $d_{y}\left(k\right)=w\sin k$.
The winding number as the topological invariant hence can be defined
in $d_{x}$-$d_{y}$ plane as
\begin{equation}
\text{Winding}=\frac{1}{2\pi i}\int_{-\pi}^{\pi}\frac{d}{dk}\log\boldsymbol{h}\left(k\right),
\end{equation}
where $\boldsymbol{h}\left(k\right)=d_{x}\left(k\right)+id_{y}\left(k\right)=v+we^{ik}$. 

Meanwhile, if $w<v$, the unit cell is $(B_{m},A_{m+1})$. Thus the bulk momentum-space Hamiltonian of the system becomes
\begin{equation}
h\left(k\right)=\left(\begin{array}{cc}
0 & w+ve^{-ik}\\
w+ve^{ik} & 0
\end{array}\right),\label{eq:Hk-1-1}
\end{equation}
and we have $\boldsymbol{h}\left(k\right)=d_{x}\left(k\right)+id_{y}\left(k\right)=w+ve^{ik}$
where $d_{x}\left(k\right)=w+v\cos k$, $d_{y}\left(k\right)=v\sin k$.

Fig.~\ref{Fig:diagram-renormal-1}(e) shows that the winding number of this SSH
model with an odd number of qubits is always 1 whether $v<w$ or $v>w$.
The wavefunctions of two edge states are presented in Fig.~\ref{Fig:diagram-renormal-1}(f).

\section{Energy spectrum detecting with Input-output theory\label{sec:Energy-spectrum-with}}

\begin{figure}[htp]
\begin{centering}
\par\end{centering}
\begin{centering}
\includegraphics[width=9cm]{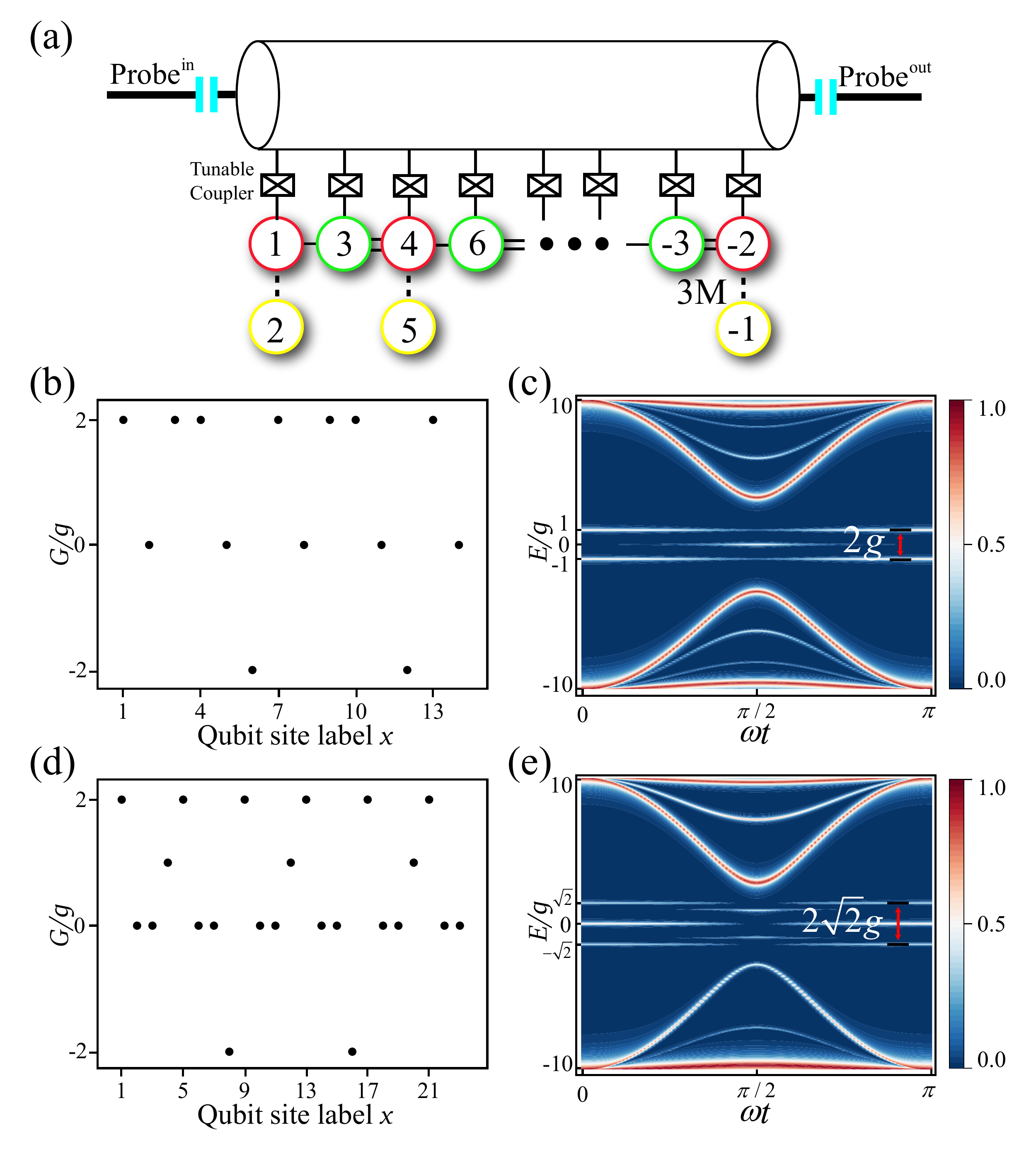}
\par\end{centering}
\caption{\label{Fig:Transmission line}(a) Schematic diagram for the SSH qubit array detected by a microwave
transmission line. The transmission line here is coupled to the
qubit chain with a set of josephson couplers, and the coupling strength between each qubit and the waveguide is tunable by the coupler. The transmission line can be seen as a long waveguide with infinite microwave modes. The frequency of input probe light is tunable and we
can get the energy spectrum of the qubit chain by detecting the transmission of the probe
light. (b) Inhomogeneous couplings between the waveguide and qubits for the extended SSH chain in Fig.~\ref{Fig:two-qubit state}(a). (c) Energy spectrum of the qubit chain for 2-qubit state transfer with waveguide-qubit couplings in (b). (d) Inhomogeneous couplings between the waveguide and qubits for the extended SSH chain in Fig.~\ref{Fig:3-qubit state transfer}(a). (e) Energy spectrum of the qubit chain for 3-qubit state transfer with waveguide-qubit couplings in (d).}
\end{figure}

To estimate the dynamical period in our QST protocol, one can measure
the eigen-energy spectrum of edge states with input-output theory for our extended SSH chain. Specifically,
for the $2$-qubit state transfer as discussed above, the energy difference between the
two edge states of the chain is $2g$. Meanwhile, for the $3$-qubit state transfer protocol, the energy difference between the
three edge states of the chain is $\sqrt{2}g$. To measure the energy spectrum of the extended SSH chain, we assume that
a weak probe field is applied to the qubit chian via a transmission
line. Thus, the energy spectrum can be detected by the transmission or reflection
of the probe light. This setup is schematically shown in Fig.~\ref{Fig:Transmission line}(a).

For the convenience of derivation, we here consider a generic qubit chain with arbitrary qubit-qubit couplings. The Hamiltonian of this qubit chain can be written as
\begin{equation}
	H_{\textrm{qubits}}=\sum_{x=1}^{\mathcal{L}}\omega_{x}\sigma_{x}^{+}\sigma_{x}^{-}+\sum_{x,y}^{x\neq y}\left(\Omega_{x,y}\sigma_{x}^{+}\sigma_{y}^{-}+\textrm{H.c.}\right)
\end{equation}
Here $\omega_{x}$ is the frequency of the $x$th qubit, and $\Omega_{x,y}$ ($x,y\in\left\{1,\cdots,\mathcal{L}\right\}$)
is the coupling strength between $x$th and $y$th qubits. When coupled with the
transmission line, the Hamiltonian of whole system will become
\begin{align}
H & =H_{\text{qubits}}+\omega_{c}a^{\dagger}a+\stackrel[x=1]{\mathcal{L}}{\sum}G_{x}\left(\sigma_{x}^{+}a+\sigma_{x}^{-}a^{\dagger}\right)\nonumber \\
 & +\varepsilon\left(ae^{i\omega_{p} t}+a^{\dagger}e^{-i\omega_{p} t}\right),
\end{align}
where $\omega_{c}$ is the frequency of waveguide mode, $G_{x}$ is the coupling
strength between waveguide and $x$th qubit, and $\varepsilon$ is the driving strength of the probe light.
After rotating-wave approximation by applying a unitary operator $U=\exp\left(\omega_{p} a^{\dagger}at+\stackrel[x=1]{\mathcal{L}}{\sum}\omega_{p}\sigma_{x}^{+}\sigma_{x}^{-}t\right)$,
we get the effective Hamiltonian as
\begin{align}
H_{eff} & =\stackrel[x=1]{\mathcal{L}}{\sum}\triangle_{x}\sigma_{x}^{+}\sigma_{x}^{-}+\stackrel[x,y]{x\neq y}{\sum}\left(\Omega_{x,y}\sigma_{x}^{+}\sigma_{y}^{-}+\text{H.c.}\right)\nonumber \\
 & +\triangle_{c}a^{\dagger}a+\stackrel[x=1]{\mathcal{L}}{\sum}G_{x}\left(\sigma_{x}^{+}a+\sigma_{x}^{-}a^{\dagger}\right)+\varepsilon\left(a+a^{\dagger}\right),
\end{align}
 where $\triangle_{x}=\omega_{x}-\omega_{p}$ and $\triangle_{c}=\omega_{c}-\omega_{p}$ are the detuning frequency
for qubits and waveguide, respectively. The waveguide naturally has infinite modes and only the mode resonating with the probe light has contribution to the Hamiltonian, i.e., $\omega_{c}=\omega_{p}$. Thus we can have $\triangle_{c}=0$. We here consider the low-excitation
limit, so the evolutions of operators' average value can be derived as
\begin{align}
\dot{\left\langle a\right\rangle } & =-\left(\kappa+i\triangle_{c}\right)\left\langle a\right\rangle -i\stackrel[x=1]{\mathcal{L}}{\sum}G_{x}\left\langle \sigma_{x}^{-}\right\rangle +\varepsilon
\end{align}
\begin{align}
\dot{\left\langle \sigma_{x}^{-}\right\rangle } & =-i\triangle_{x}\left\langle \sigma_{x}^{-}\right\rangle -iG_{x}\left\langle a\right\rangle -i\stackrel[y=1]{\mathcal{L}}{\sum}\Omega_{x,y}\left\langle \sigma_{y}^{-}\right\rangle \nonumber \\
 & -\Gamma_{x}\left\langle \sigma_{x}^{-}\right\rangle,
\end{align}
where $\kappa$ is decay rate for the waveguide and $\Gamma_{x}$ is the decay of $x$th qubit. The evolution equations above can be written
into matrix form as
\begin{eqnarray}
\dot{\left\langle a\right\rangle } & = & -\left(\kappa+i\triangle_{c}\right)\left\langle a\right\rangle -i\boldsymbol{G}^{T}\boldsymbol{\sigma}+\varepsilon\\
\dot{\left\langle \boldsymbol{\sigma}\right\rangle } & = & -i\left(\boldsymbol{\triangle}+\boldsymbol{\Omega}-i\boldsymbol{\Gamma}\right)\left\langle \boldsymbol{\sigma}\right\rangle -i\boldsymbol{G}\left\langle a\right\rangle,
\end{eqnarray}
where $\boldsymbol{\sigma}=\left(\left\langle \sigma_{1}^{-}\right\rangle ,\left\langle \sigma_{2}^{-}\right\rangle, \cdots\left\langle \sigma_{\mathcal{L}}^{-}\right\rangle \right)^{T}$,
$\boldsymbol{\triangle}=\textrm{Diag}\left(\triangle_{1},\triangle_{2},\cdots,\triangle_{\mathcal{L}}\right)$, $\boldsymbol{\Gamma}=\textrm{Diag}\left(\Gamma_{1},\Gamma_{2},\cdots,\Gamma_{\mathcal{L}}\right)$ ,
$\boldsymbol{G}=\left(G_{1},G_{2},\cdots,G_{\mathcal{L}}\right)^{T}$ and $\boldsymbol{\Omega}$ has the matrix form of the coupling terms in $H_{\text{qubits}}$, i.e., $\boldsymbol{\Omega}_{x,y}=\Omega_{x,y}$. Here, notice that for our measurement setup as shown in Fig.~\ref{Fig:Transmission line}(a), the transmission line only couples to the top half of the chain. This feature can be reflected by the coupling coefficient $G_{x}$ as shown in Fig.~\ref{Fig:Transmission line}(b) and (d).

When the system is stable, i.e., $\dot{\left\langle a\right\rangle }=\dot{\boldsymbol{\sigma}}=0$,
we can get the steady solution of the equations as
\begin{align}
\boldsymbol{\sigma} & =-i\boldsymbol{M}^{-1}\boldsymbol{G}\left\langle a\right\rangle
\end{align}
\begin{align}
\left\langle a\right\rangle  & =\frac{\varepsilon}{\kappa+i\triangle_{c}+\boldsymbol{G}^{T}\boldsymbol{M}^{-1}\boldsymbol{G}}
\end{align}
where $\boldsymbol{M}=i\boldsymbol{\triangle}+i\boldsymbol{\Omega}+\boldsymbol{\Gamma}$.
From the input-output theory, the transmission of the probe field is
\begin{align}
t_{p} & =1-\frac{\kappa}{\varepsilon}\left\langle a\right\rangle \nonumber \\
 & =1-\frac{\kappa}{\kappa+i\triangle_{c}+\boldsymbol{G}^{T}\boldsymbol{M}^{-1}\boldsymbol{G}},  \label{transmission}
\end{align}
and therefore the reflection of the probe field is
\begin{equation}
r_{p}=\frac{\kappa}{\kappa+i\triangle_{c}+\boldsymbol{G}^{T}\boldsymbol{M}^{-1}\boldsymbol{G}}.\label{reflection}
\end{equation}

For the 2-qubit state transfer protocol with 14 qubits chain in Fig.~\ref{Fig:two-qubit state}(a), all the qubits are resonating with each other, i.e., $\omega_{x}=\omega_{q}$ for $x=1,2,...,\mathcal{L}$. Therefore we have $\boldsymbol{\triangle}=\textrm{Diag}\left(\triangle_{q},\triangle_{q},\triangle_{q},\cdots,\triangle_{q}\right)$ and $\triangle_{q}=\omega_{q}-\omega_{p}$. $\boldsymbol{\Omega}$ is the matrix form of the Hamiltonian shown in Eq.~(\ref{eq:Hamiltonian-2}). The decay of the waveguide is $\kappa=2.5g$ and all the qubits have the same decay rate as $\gamma=0.01g$. The coupling strengths $\boldsymbol{G}$ between qubits and waveguide are shown in Fig.~\ref{Fig:Transmission line}(b). From Eq.~(\ref{reflection}) we can get the reflection spectrum of the qubit chain for different probe light as shown in Fig.~\ref{Fig:Transmission line}(c), where the vertical coordinates represent the detuning frequency $\triangle_{q}$ and also the eigen-energy of the system. Two edge states are clearly shown in the figure and we can get the energy shift as $2g$. Another case we show here is the 23 qubits chain for 3-qubit state transfer, and we can also get the energy shift as $2\sqrt{2}g$ in Fig.~\ref{Fig:Transmission line}(e).

\end{document}